\documentclass[fleqn,usenatbib]{mnras}

\usepackage{newtxtext,newtxmath}

\usepackage[T1]{fontenc}

\DeclareRobustCommand{\VAN}[3]{#2}
\let\VANthebibliography\thebibliography
\def\thebibliography{\DeclareRobustCommand{\VAN}[3]{##3}\VANthebibliography}

\usepackage{graphicx}	
\usepackage{amsmath}	

\usepackage[dvipsnames,svgnames,table,rgb]{xcolor}
\usepackage{caption}

\newcommand{\Msun}{\mathrm{M}_{\odot}}
\newcommand{\kms}{\rm km\,s^{-1}}

\title[NGC~1916: NSC of the LMC]{NGC~1916: the Nuclear Star Cluster of the Large Magellanic Cloud}

\author[E. A. Balakina et al.]{
Elena A. Balakina,$^{1}$\thanks{E-mail: E.Balakina@2022.ljmu.ac.uk (EB)}
Sebastian Kamann,$^{1}$\thanks{E-mail: S.Kamann@ljmu.ac.uk (SK)}
Anil Seth$^{2}$\thanks{E-mail: aseth@astro.utah.edu (AS)},
Florian Niederhofer$^{3}$,
Nate Bastian$^{4,5}$,\newauthor
Ivan Cabrera-Ziri$^{6}$,
Nadine Neumayer$^{7}$,
Julia Bodensteiner$^{8}$,
Stefan Dreizler$^{9}$,
L. Clifton Johnson$^{10}$,\newauthor
Sara Saracino$^{11}$
\\
$^{1}$Astrophysics Research Institute, Liverpool John Moores University, IC2 Liverpool Science Park, 146 Brownlow Hill, Liverpool L3~5RF, UK\\
$^{2}$Department of Physics and Astronomy, University of Utah, Salt Lake City, UT 84112, USA\\
$^{3}$Leibniz-Institute for Astrophysics (AIP), An der Sternwarte 16, 14482 Potsdam, Germany\\
$^{4}$Donostia International Physics Center (DIPC), Paseo Manuel de Lardizabal, 4, 20018, Donostia-San Sebastián, Guipuzkoa, Spain\\
$^{5}$IKERBASQUE, Basque Foundation for Science, 48013, Bilbao, Spain\\
$^{6}$Vyoma GmbH, Karl-Theodor-Stra\ss e 55, 80803 Munich, Germany \\
$^{7}$Max Planck Institute for Astronomy, K\"onigstuhl 17, D-69117 Heidelberg, Germany \\
$^{8}$Anton Pannekoek Institute for Astronomy, University of Amsterdam, 1090 GE Amsterdam, The Netherlands \\
$^{9}$Institut f\"ur Astrophysik und Geophysik (IAG), Georg-August-Universit\"at G\"ottingen, Friedrich-Hund-Platz 1, 37077 G\"ottingen, Germany \\
$^{10}$Center for Interdisciplinary Exploration and Research in Astrophysics (CIERA), Northwestern University, 1800 Sherman Ave, Evanston, IL 60201, USA \\
$^{11}$INAF – Osservatorio Astrofisico di Arcetri, Largo E. Fermi 5, 50125 Firenze, Italy \\
}

\date{Accepted XXX. Received YYY; in original form ZZZ}

\pubyear{\the\year{}}

\begin{document}
\label{firstpage}
\pagerange{\pageref{firstpage}--\pageref{lastpage}}
\maketitle

\begin{abstract}

Nuclear star clusters are found at the centres of nearly all galaxies with masses between 0.1 and 10 billion solar masses. The Large Magellanic Cloud is the nearest galaxy in this mass range, yet its distorted morphology has made it difficult to confirm whether it hosts a nuclear star cluster. One of the most promising candidates is NGC~1916, which we examine in this work.
Using VLT/MUSE spectroscopic observations, we analyse the kinematics and metallicities of its brightest stars and combine these with colour-magnitude diagram information from the Hubble Space Telescope. We use the kinematics and number density profile to distinguish member and non-member stars. 
The member stars along the red giant branch show a strong bimodality in their measured metallicities, with distinct peaks at [M/H] of $-1.6$ and $-0.5$. Our analysis also reveals a population of young massive main-sequence stars inside the cluster, including numerous Be stars with emission lines.
The spatial distributions of the three components differ, with the young main-sequence stars showing the highest concentration and central dispersion. The cluster appears to have formed stars over a long period of time with the young and metal-rich populations spanning ages from $\sim$100~Myr to at least 1~Gyr, while the metal-poor population is consistent with a population with an age of $\sim$12~Gyr.
This extended star formation history and diverse stellar populations are unique among clusters in the Large Magellanic Cloud and strongly indicate that NGC~1916 is the nuclear star cluster of the galaxy. We examine the cluster properties in the context of other nuclear star clusters and discuss the implications for the possible presence of a massive black hole in the Large Magellanic Cloud.

\end{abstract}




\section{Introduction}
\label{sec:intro}

The Large Magellanic Cloud (LMC) is the nearest late-type dwarf galaxy in the Local Group, located at a distance of approximately $50$~kpc. With a stellar mass of $2.5\times10^9~\Msun$, it is one of the most massive satellites in the Local Group~\citep{2012AJ....144....4M}.

Numerous observations have revealed the LMC's complex morphology as a disc galaxy \citep[e.g.,][]{2001AJ....122.1807V, 2002AJ....124.2639V} undergoing perturbations caused by interactions with its closest neighbour, the Small Magellanic Cloud \citep[SMC;][]{2005MNRAS.356..680B, 2011ApJ...737...29O, 2013ApJ...764..161K, 2018ApJ...866...90C}. Furthermore, the LMC-SMC system as a whole is currently infalling towards the Milky Way \citep{2005A&A...432...45B}, which is also shaping the LMC structure \citep{2019MNRAS.482L...9B, 2019MNRAS.488..918T} and resulting in star formation activity \citep{2009AJ....138.1243H, 2021MNRAS.508..245M, 2022MNRAS.513L..40M}. Nevertheless, as the LMC is likely on its first passage around the Milky Way \citep{2007ApJ...668..949B, 2012MNRAS.421.2109B}, it was not significantly affected by it until recently, and thus should have had a history similar to that of other late-type galaxies of comparable mass.

\begin{figure}
    \centering
	\includegraphics[width=1.\columnwidth]{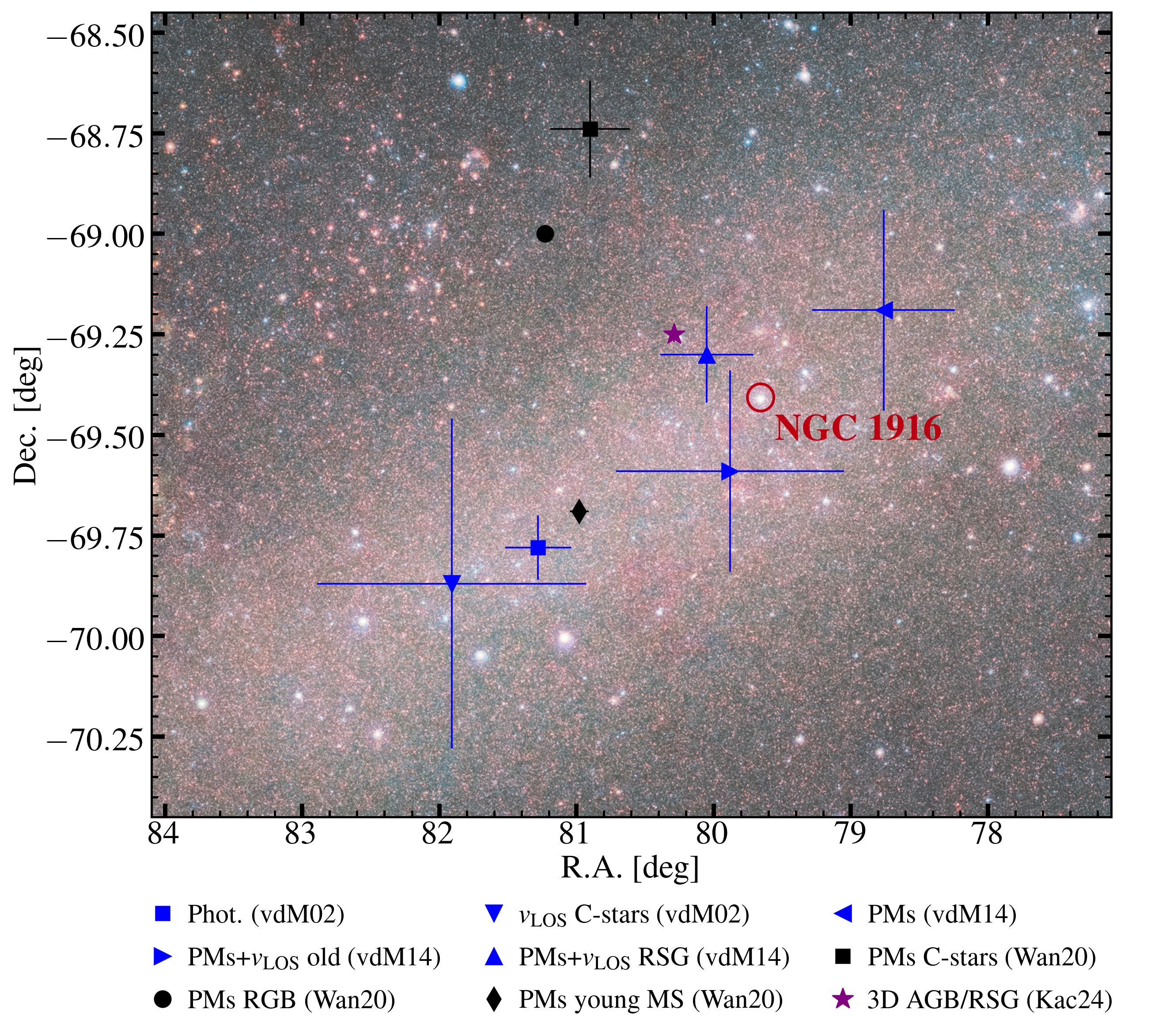}
    \caption{Central $2.45^\circ \times 2^\circ$ region of the LMC from the ESO/VMC Survey. The open red circle outlines the region within the tidal radius of NGC~1916. The blue square marks the photometric centre of the LMC bar, and the downward-pointing blue triangle marks the centre inferred from the line-of-sight velocities of carbon stars \citep{2002AJ....124.2639V}. The remaining blue triangles show the centres derived by \citet{2014ApJ...781..121V} from HST proper motions alone (left-pointing), or in combination with line-of-sight velocities of old stars (right-pointing) or red supergiants (upward-pointing). The black symbols show the centres derived by \citet{2020MNRAS.492..782W} from \textit{Gaia} DR2 PMs of carbon stars (square), red giant branch stars (circle), and young main-sequence stars (diamond). The purple star marks the centre derived by \citet{2024AA...692A..40K} from the three-dimensional kinematics of asymptotic giant branch stars and red supergiants in \textit{Gaia} DR3. Error bars show the reported uncertainties in right ascension and declination. The figure illustrates the scatter among the LMC centres derived using different methods, with an RMS distance of about 415~pc from their mean position. The full comparison is given in Table~\ref{tab:lmc_centres}.}
    \label{fig:centres}
\end{figure}

A direct consequence of these interactions is that the characterisation of the centre of the LMC, both in terms of its location and its nature, remains uncertain. Estimates of the central coordinates inferred from stellar dynamics differ by up to 1~kpc, see Fig.~\ref{fig:centres}, depending on the adopted methodologies and the types of tracers used \citep[e.g.][]{2014ApJ...781..121V, 2020MNRAS.492..782W, 2021AA...649A...7G, 2022MNRAS.512.5423N, 2024AA...692A..40K, 2025AA...700A.279V}. At the relatively close distance of the LMC, such differences correspond to projected offsets of around 1~degree, hampering observational efforts to determine whether the LMC hosts a central massive black hole (MBH), a nuclear star cluster (NSC), or both.

Dense and massive NSCs are nearly ubiquitous at the centres of galaxies within the LMC mass range: NSCs are found in $\sim$80\% of galaxies with stellar masses of $10^8$--$10^{10}~\Msun$ \citep{2019ApJ...878...18S, 2021MNRAS.507.3246H, 2023ApJ...958..100A}, making it likely that the LMC hosts one. If present, an NSC would serve as a useful tracer of its host galaxy's dynamical centre. Among all clusters identified near the various proposed LMC centres, the most promising NSC candidate is the globular cluster NGC~1916. With a mass of $6 \times 10^5~\Msun$ \citep{2003MNRAS.338...85M}, it falls within the NSC mass range observed in galaxies of comparable stellar mass \citep{2020A&ARv..28....4N}, and it is located within the most central part of the LMC, as shown by comparison with all proposed centres in Fig.~\ref{fig:centres}. One might expect that tidal interactions could displace a potential LMC NSC from the dynamical centre. However, \citet{2020MNRAS.492..782W} showed that young stars are offset by only about $1^\circ$ from old stars, reflecting the relatively mild impact of the LMC--SMC interaction on the central region. It is therefore reasonable to expect that an NSC would remain within this region.

As opposed to globular clusters in galactic fields, NSCs frequently exhibit complex stellar populations with wide metallicity and/or age spreads, indicative of extended star formation histories \citep[see][for a review]{2020A&ARv..28....4N}. Two main scenarios have been proposed for the formation of NSCs:
\begin{enumerate}
    \item \textit{Ex-situ} formation, in which pre-existing globular clusters sink towards the galactic centre via dynamical friction \citep{1975ApJ...196..407T} and coalesce to form a central overdensity. Simulations show that disrupted globular clusters can supply most of the NSC mass in galaxies with stellar masses up to $10^{11}~\Msun$ \citep{2014ApJ...785...71G}. This channel is expected to be more important in low-mass galaxies, where NSCs tend to be relatively metal-poor and chemically similar to globular clusters \citep{2021A&A...650A.137F, 2022A&A...658A.172F}.
    \item \textit{In-situ} formation, in which stars form directly within a galactic centre. This channel may range from early formation with limited gas supply, producing a relatively simple chemical composition, to repeated star formation episodes sustained by gas inflows. The latter can produce young stellar populations and metallicities exceeding those of the surrounding galaxy \citep{2014ApJ...785...71G}. 
\end{enumerate}
These two channels are not mutually exclusive. A combination of \textit{in-situ} and \textit{ex-situ} mechanisms may operate simultaneously, with gas accretion occurring alongside the inspiral of old star clusters, or with infalling clusters themselves being gas-rich and continuing to form stars during their migration. Given this diversity of formation pathways, recent studies have focused on distinguishing between channels and assessing how host galaxy properties and large-scale environment influence NSC evolution, from both observational \citep[e.g.,][]{2018MNRAS.480.1973K, 2021A&A...650A.137F, 2022A&A...658A.172F, 2022A&A...667A.101F} and theoretical \citep[e.g.,][]{2012ApJ...750..111A, 2014ApJ...785...71G, 2016MNRAS.461.3620G, 2022MNRAS.516.4691L, 2025ApJ...979...85L} perspectives. 

Both NSCs and massive black holes co-exist at the centres of many present-day galaxies and stripped NSCs \citep[e.g.,][]{2008ApJ...678..116S, 2009MNRAS.397.2148G, 2020A&ARv..28....4N, 2024Natur.631..285H}. Some models suggest that these two types of objects are directly linked, with dense NSCs potentially providing a seeding mechanism for massive black holes through the formation of supermassive stars or successive mergers of stellar-mass black holes \citep[e.g.,][]{2004Natur.428..724P, 2024MNRAS.531.3770R, 2026arXiv260107917R, 2026arXiv260204176N}; see \citet{2020ARA&A..58..257G} for a review. Owing to their very high stellar densities, NSCs are also able to survive the accretion of their host galaxies onto more massive systems \citep[e.g.,][]{2013MNRAS.433.1997P}. At the high-mass end, stripped NSCs -- also known as ultra-compact dwarf galaxies (UCDs) -- have been found to host supermassive black holes above $10^6~{\rm \Msun}$ \citep{2014Natur.513..398S, Ahn2018, Afanasiev2018}.

Within the Milky Way, several clusters are believed to be stripped NSCs \citep[see][for an overview]{2021MNRAS.500.2514P}, with the most prominent examples being M~54 and $\omega$~Cen. The detailed properties available for these nearby systems are particularly valuable for constraining formation scenarios. Several studies \citep{2007ApJ...667L..57S, 2019ApJ...886...57A, 2019MNRAS.490L..67S} have shown that M~54 contains at least two distinct stellar populations with significant metallicity and age differences, where the old, metal-poor population was likely formed \textit{ex-situ}, while younger stars formed later through gas accretion, i.e., \textit{in-situ}. On the other hand, $\omega$~Cen exhibits an even more complex chemical composition \citep[e.g.,][]{2017ApJ...844..164B} but shows considerably less age spread \citep{2024ApJ...977...14C}. This is likely a consequence of the earlier accretion of its host: while the origin of $\omega$~Cen is still debated, M~54 can securely be linked to the recent accretion of the Sagittarius (Sgr) dwarf, owing to its location in the Sgr stream.

In this context, the putative NSC of the LMC is of particular interest. Additional motivation comes from the possible presence of an MBH within the galaxy. Kinematic analyses of \textit{Gaia} DR2 \citep{2019MNRAS.483.2007E} and DR3 \citep{2025ApJ...982..188H} suggest that approximately half of the known hyper-velocity stars (HVSs) originate from the LMC. The extreme velocities of these stars are believed to result from interactions between binary systems and a massive black hole \citep{1988Natur.331..687H}, pointing to an MBH of $10^5$--$10^6~\Msun$ within the LMC. Independently, \citet{2017ApJ...846...14B} used a wide yet shallow mosaic of MUSE observations to search for such an object and placed an upper limit of $10^{7.1}~\Msun$ on its mass. Given that the coexistence of MBHs and NSCs is commonly observed in galactic centres \citep{2020A&ARv..28....4N}, NGC~1916 may represent the most natural site in which to search for a central MBH in the LMC. Spectroscopic measurements have found the cluster to be metal-poor, with metallicities ranging between [Fe/H] of -2.08 and -1.54 \citep{Olszewski1991, Colucci2012}, and highly alpha-enhanced with [Si/Fe]$=$0.39 \citep{Mucciarelli2021}. NGC~1916's colour-magnitude diagram shows that it hosts an old population. However, differential reddening has limited its interpretation \citep{1998MNRAS.300..665O}.

\begin{figure*}
    \centering
	\includegraphics[width=1.\textwidth]{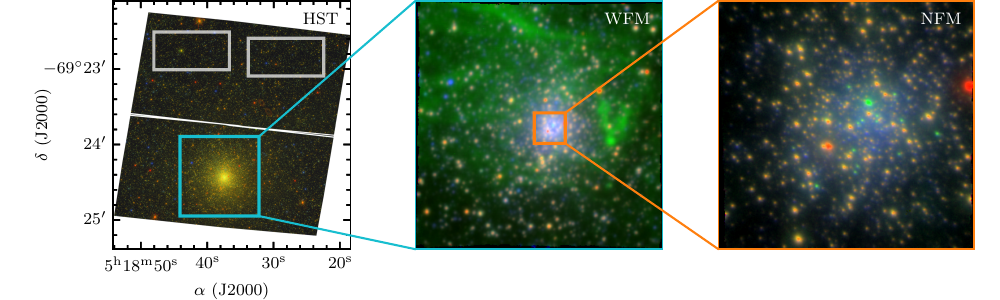}
    \caption{The HST data of NGC~1916 is shown as a F300X/F606W/F814W colour composite in the left panel. Two white rectangles indicate the $1 \times 0.5$~arcmin$^2$ patches used to sample the field population. The location of the MUSE WFM pointing is indicated by a cyan square, while a colour composite generated from the WFM data is shown in the middle panel. The red and blue bands were obtained by integrating the data over the $z$ and $g$ throughput curves, respectively, while the green band was generated using an H$\alpha$ narrowband filter. An orange square indicates the location of the MUSE NFM pointing, shown in the right panel using the same colour bands as for the WFM data.}
    \label{fig:footprint}
\end{figure*}

This paper is organised as follows. Section~\ref{sec:data} presents the photometric and spectroscopic data used in the analysis. Section~\ref{sec:analysis} examines the metallicities, kinematics, and radial distributions of the cluster populations, following a careful removal of field stars. The dynamical mass estimate and probable star formation history of NGC~1916 are discussed in Section~\ref{sec:discussion}, and our conclusions are summarised in Section~\ref{sec:conclusions}.

\section{DATA}
\label{sec:data}
Earlier studies of this object \citep[e.g.,][]{1998MNRAS.300..665O} were hampered by the strong differential reddening observed across the face of NGC~1916. In this study, we aim to overcome this limitation through a combination of \textit{Hubble} Space Telescope (HST) photometry covering the cluster and surrounding field; MUSE integral field spectroscopy of the central region at high spatial resolution; and \textit{Gaia} DR3 proper motions of the brightest stars in the cluster outskirts and the LMC field.
\subsection{HST}
\label{sec:hst}

We used archival HST WFC3 data from programme GO-16361 (PI: Ferraro) in the F300X, F606W, and F814W passbands~\citep{2020hst..prop16361F}. We analysed the data using \textsc{Dolphot} \citep{2016ascl.soft08013D} and produced photometric and astrometric catalogues in all three available passbands. In the left panel of Fig.~\ref{fig:footprint}, we show the footprint of the HST data, with the cluster visible on the lower (southern) of the two WFC3 chips.

We estimate the field contamination by comparing star counts across different regions. Figure~\ref{fig:footprint} shows the cluster region (cyan square), chosen to match the MUSE field of view of $1 \times 1$~arcmin$^2$, along with two field regions of $1 \times 0.5$~arcmin$^2$ each (white rectangles). The field regions were positioned on opposite sides of the HST field of view, sufficiently distant from the cluster centre to minimise its contribution while accounting for potential non-uniformity in the field star distribution. We note that different field placement resulted in variation consistent with Poisson count differences.

Within the central $1 \times 1$~arcmin$^2$ of the cluster area, we count $N_{\rm cluster} = 1086$ bright stars ($\rm F814W < 19^m$). The two background fields contain $N^1_{\rm field} = 95$ and $N^2_{\rm field} = 112$ stars above this threshold, respectively. Scaling to the same $1 \times 1$~arcmin$^2$ area yields an estimated field contribution of $N_{\rm field} = 207 \pm 14$, corresponding to a fractional contamination over the full field of view of the MUSE data of $\sim$19\%. We discuss our approach to separating cluster and field stars in Section~\ref{sec:kinematics} below.

We show the CMD of the central cluster region and the associated background regions in Fig.~\ref{fig:cmd1}. The central field has a number of features that are both shared (red clump, bright young stars) and different from the background. The cluster CMD shows thicker sequences than the field, suggesting differential extinction, which complicates its interpretation \citep{1998MNRAS.300..665O}. While we do use the cluster colour-magnitude diagram in this paper, we prioritize information from our MUSE spectra to understand the cluster stellar populations. 


\subsection{MUSE}
\label{sec:muse}

Besides the HST photometry, we analysed the spectroscopic data obtained with the Multi Unit Spectroscopic Explorer (MUSE, \citealt{2010SPIE.7735E..08B}) in program 108.2258 (PI: Kamann). Observations were obtained for one Wide Field Mode (WFM) pointing covering 1$\times$1~arcmin$^2$ and one Narrow Field Mode (NFM) pointing covering $7.5 \times 7.5$~arcsec$^2$. As shown in Fig.~\ref{fig:footprint}, both pointings were centred on NGC~1916. During the WFM observations, ground-layer adaptive optics were used, resulting in a spatial resolution of $\sim$0.7~arcsec FWHM. The NFM data, on the other hand, have HST-like spatial resolution ($<$0.1~arcsec), thanks to the use of laser-tomography adaptive optics. The WFM data, consisting of six exposures of 600~seconds each, were observed during the night of 2021-10-05, while the NFM data, consisting of four exposures of 600~s each, were obtained on 2021-12-09.

All collected data were reduced using the official MUSE pipeline presented in \citet{2020A&A...641A..28W}. The pipeline performs the basic reduction steps (bias removal, signal tracing, flat-fielding and wavelength calibration) separately for the data collected by each of the 24 MUSE spectrographs. Afterward, the data from all spectrographs are combined and resampled to produce a regular three-dimensional data cube. This step also includes sky subtraction. In the last step of the reduction, the different exposures from each instrument configuration (WFM or NFM) were combined to produce the final two data cubes, which are analysed below.

In Fig.~\ref{fig:footprint}, we show RGB colour images of both cubes. The red and blue channel images were created by integrating the MUSE cubes over the SDSS~$z$- and $g$-bands, respectively, while the green channel images were obtained using a narrow-band integration around H$\alpha$. The distinct green hue visible in the WFM image (central panel of Fig.~\ref{fig:footprint}) shows the presence of extended ionised gas, with a cavity that coincides with the cluster centre. In the NFM image shown in the right panel of Fig.~\ref{fig:footprint}, several green dots highlight the presence of numerous stars with H$\alpha$ emission within a few arcseconds of the cluster centre. We will discuss these features further in the subsequent chapters.

\begin{figure}
    \centering
	\includegraphics[width=1\columnwidth]{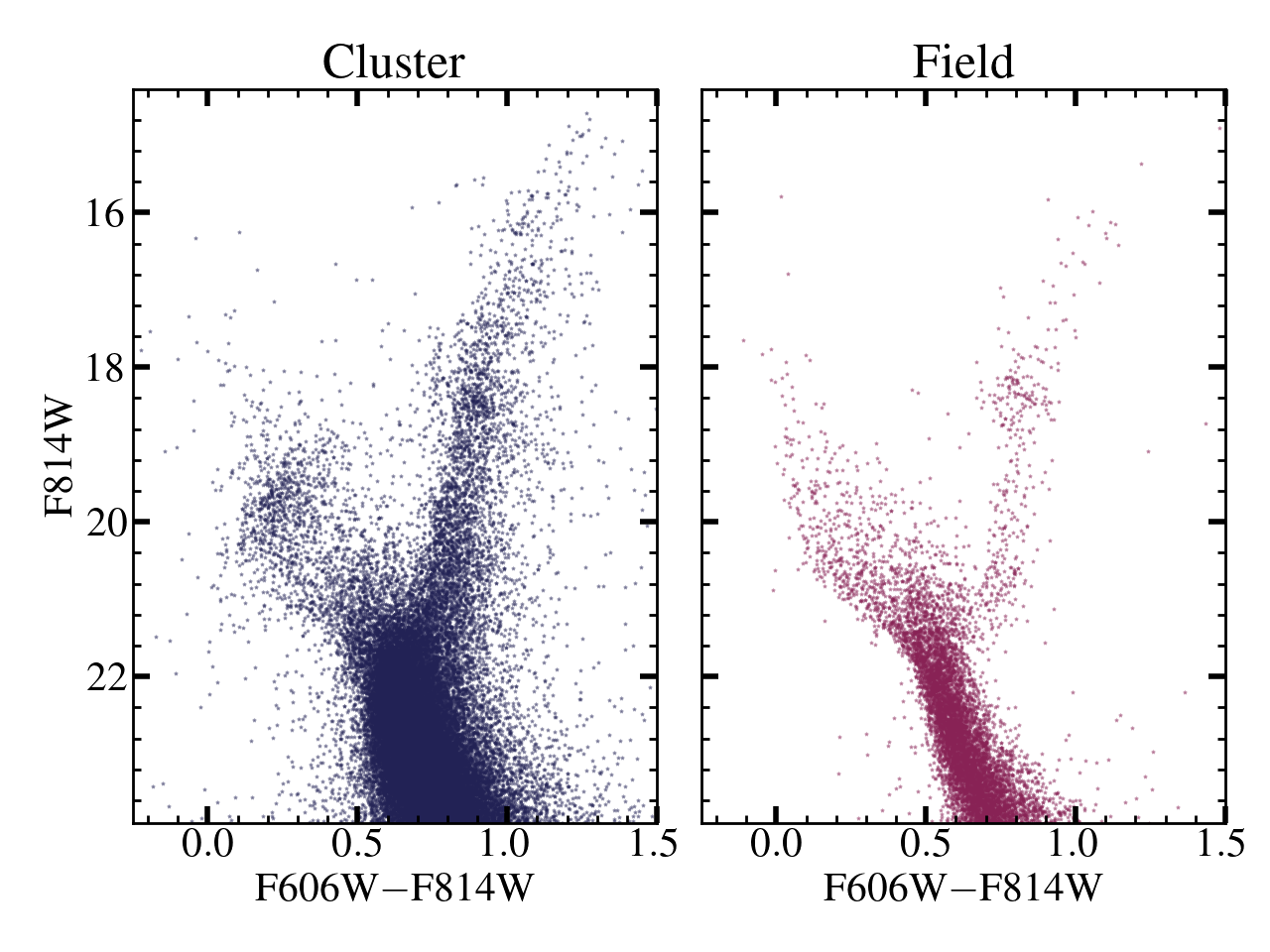}
    \caption{\textit{Left:} HST CMD of NGC~1916 within $1 \times 1$~arcmin$^2$ around the cluster centre (cyan square on Fig.~\ref{fig:footprint}). \textit{Right:} the field CMD within the rectangles shown in white on Fig.~\ref{fig:footprint}, which cover the same area as the cyan square used to construct the CMD shown in the left panel.}
    \label{fig:cmd1}
\end{figure}

To obtain stellar spectra from the MUSE cubes, we used \textsc{PampelMuse} \citep{2013A&A...549A..71K}, a code for deblending spectra from integral-field spectroscopy data via point spread function (PSF) fitting. \textsc{PampelMuse} uses an astrometric reference catalogue as input and then iteratively determines the source coordinates and the PSF shape in the cube as a function of wavelength. In the last step, this information is used to optimally extract the spectra of resolved sources from the data. As a reference catalogue, we used the HST data presented in Sec.~\ref{sec:hst} above. When analysing the WFM data, the PSF was modelled as a spherical Moffat profile, whereas we used the \texttt{MAOPPY} model developed by \citet{2019A&A...628A..99F} whilst analysing the NFM data.

The extracted spectra were analysed using \textsc{Spexxy}\footnote{\url{https://github.com/thusser/spexxy}}~\citep{2016A&A...588A.148H}, a code to measure line-of-sight velocities ($v_{\rm LOS}$) and stellar parameters via full-spectrum fitting against a library of template spectra. Here, we adopted different approaches to analyse the spectra of cool and hot stars.
\vspace{-\topsep}
\begin{itemize}
\item For cool stars, hereafter defined as those with a photometric colour of F606W-F814W~$>0.6$, we used the \texttt{GLib} spectra published by \citet{2013A&A...553A...6H} and constrained $v_{\rm LOS}$, the scaled-solar metallicity ${\rm [M/H]}$, and the effective temperature $T_{\rm eff}$ for each extracted spectrum. The surface gravities of the stars were instead measured by comparing the HST optical CMD with an isochrone from the \textit{MIST} database \citep{2016ApJS..222....8D, 2016ApJ...823..102C} and selecting the nearest points in CMD space. To generate the isochrone, we assumed a metallicity of the cluster of ${\rm [M/H]}=-1.5$, an age of 12.59~Gyr, and an extinction of $A_{\rm V}=0.58$~mag. Metallicity and extinction were adopted from Table~1 of \citet{2019MNRAS.482.1275U}, whereas the age was inferred by visual inspection of the isochrone fits. We note that we used the same isochrone to obtain initial guesses for the effective temperatures in the \textsc{Spexxy} fits. The $v_{\rm LOS}$ initial guesses were instead obtained by cross-correlating the spectra with templates from the \texttt{GLib} library with matching parameters (based on the isochrone comparison).

The [M/H] measurements of the cool stars are a key part of our analysis. To double-check the reliability of our results from full-spectrum fitting, we also calculated metallicities based on just the \ion{Ca}{II} triplet at 8498, 8542, and 8662~{\text\AA} \citep[e.g.,][]{2013MNRAS.434.1681C, 2019ApJ...886...57A}. The full analysis is presented in Appendix~\ref{sec:metal}; while we see some offsets between the results, the bimodal feature discussed in Sec.~{\ref{sec:me}} below is clearly seen in the \ion{Ca}{II} triplet metallicities as well.
\item Hot stars, i.e., those bluer than F606W-F814W=0.6, were instead fitted against the library published by \citet{2018A&A...618A..25A}, which extends to higher effective temperatures than \texttt{GLib} ($T_{\rm eff}\leq30~{\rm kK}$ instead of $\leq15~{\rm kK}$). At the brightness level of our study (F814W~$\leq19$, see below), which is brighter than the extended horizontal branch of NGC~1916 (cf. Fig.~\ref{fig:cmd1}), the hot stars are expected to be predominantly young massive main-sequence stars. Therefore, we initialised the fits with a common $\log g=4$ and an effective temperature of $T_{\rm eff}=10~{\rm kK}$. As the MUSE spectra of such stars lack any metal lines, we fixed the metallicity to [M/H]=-0.5, as expected for young LMC stars. Thus, three parameters were fitted: $\log g$, $T_{\rm eff}$, and $v_{\rm LOS}$. The initial guess for the latter was $265~{\rm km\,s^{-1}}$, roughly the systemic velocity of both NGC~1916 and the surrounding LMC. The right panel of Fig.~\ref{fig:footprint} suggests the presence of stars with H$\alpha$ emission, hence we visually inspected the spectra prior to the analysis and masked out the H$\alpha$ line in cases were emission was present.
\end{itemize}
\vspace{-\topsep}

We applied several quality cuts to ensure the robustness of the following analyses. Specifically, we only kept spectra with a signal-to-noise ratio (SNR) $>15$ for which \textsc{PampelMuse} determined a \texttt{MagAccuracy}$~>0.6$. \texttt{MagAccuracy} quantifies the agreement between the magnitudes recovered from the MUSE spectra and those available in the reference catalogue and ranges from 0 (poor agreement) to 1 (close match). Only spectra with a formally successful \textsc{Spexxy} analysis were retained. For an easier assessment of the expected field contamination in our sample, we further discarded any results from spectra with magnitudes F814W~$>19$. Finally, for the cool subsample (F606W-F814W$>0.6$), we enforced an uncertainty on [M/H] of $\leq0.1$~dex and a velocity accuracy of better than $5~{\rm km\,s^{-1}}$. These cuts were not applied to the hot subsample; while [M/H] was kept fixed when fitting the spectra, the line-of-sight velocities tend to be less accurate given the absence of narrow absorption lines. We verified that all hot stars entering our analysis had velocity uncertainties $<10~{\rm km\,s^{-1}}$. If the same star is observed with both WFM and NFM, we keep only the NFM measurement to avoid duplicates and because the NFM spectra have higher S/N for all stars present in both subsets. 

Our final MUSE sample consists of $606$ stars observed in the WFM and $299$ in the NFM ($905$ in total). Comparing these with the HST samples of stars brighter than F814W=19 within the respective footprints (defined as the convex hulls around our final NFM and WFM samples) yields estimates of the overall MUSE completeness of $67$\% (606/901) for the WFM (excluding the NFM footprint) and $68$\% (299/441) for the NFM. However, as will be discussed in Sec.~\ref{sec:memb} below, the completeness varies strongly with distance to the cluster centre. The cool subsample comprises 806 stars in total (of which 237 have NFM spectra), while the hot subsample includes 99 stars (NFM: 62).

\subsection{Gaia DR3}
\label{sec:gaia}
We supplemented the spectroscopic dataset with proper motion measurements from the third data release of the \textit{Gaia} mission \citep[DR3;][]{2023A&A...674A...1G}, which provides proper motions for over $1.4$ billion sources. We cross-matched the \textit{Gaia} DR3 catalogue with our HST photometric catalogue. To ensure reliable astrometric solutions, we applied the following quality cuts to the matched \textit{Gaia} sample: a renormalised unit weight error $\mathrm{RUWE} < 1.4$, an astrometric excess noise $< 0.9$\,mas, a minimum fraction of good along-scan observations $> 0.02$, and proper motion uncertainties in both right ascension and declination $< 0.5$\,mas\,yr$^{-1}$. The resulting sample contains $591$ stars within the HST field of view, of which $65$ also have a MUSE counterpart. 

\section{Analysis}
\label{sec:analysis}

\subsection{Metallicity Bimodality}
\label{sec:me}

\begin{figure}
    \centering
    \begin{minipage}[t]{0.49\columnwidth}
        \centering
        \includegraphics[width=\linewidth]{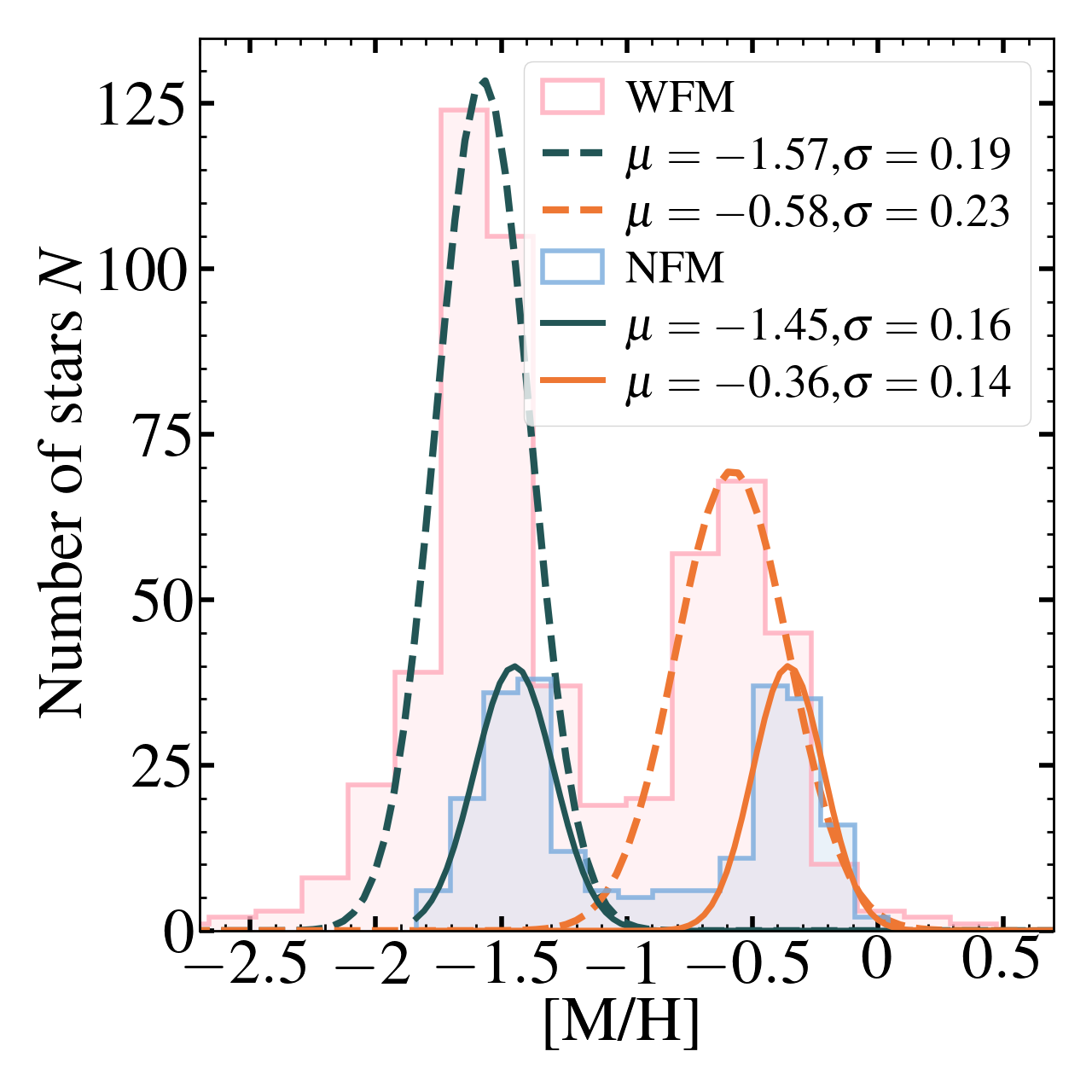}
    \end{minipage}\hfill
    \begin{minipage}[t]{0.49\columnwidth}
        \centering
        \includegraphics[width=\linewidth]{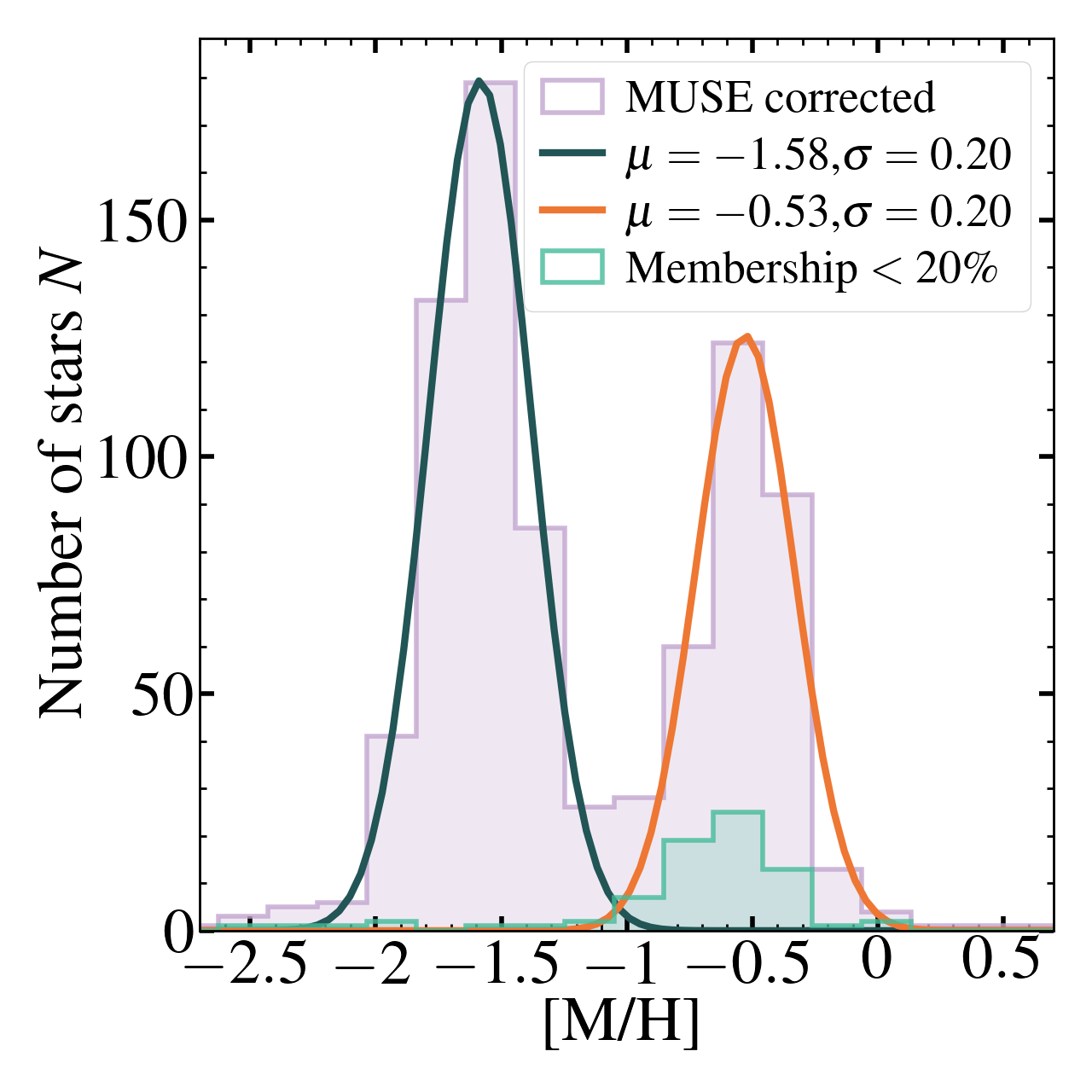}
    \end{minipage}
    \caption{\textit{Left:} NGC~1916 metallicity histograms obtained from MUSE wide-field mode (WFM; pink) and narrow-field mode (NFM; blue) data of the cool MUSE subsample. Dashed and solid curves show the Gaussian components fitted to the WFM and NFM samples, respectively. In both panels, dark cyan shows the metal-poor component and orange the metal-rich component. \textit{Right:} Combined MUSE metallicity distribution (purple) after subtracting $0.125$~dex from all NFM metallicities to align the NFM and WFM metal-poor peaks. The mint coloured histogram shows the stars with membership probabilities below $20\%$, see Sec.~\ref{sec:kinematics}.}
    \label{fig:me_hist_corrected}
\end{figure}

The left panel of Fig.~\ref{fig:me_hist_corrected} shows the metallicity distributions of the WFM and NFM subsamples of cool stars. We observe a clear bimodality, with a metal-poor peak near $[\mathrm{M/H}]=-1.5$ and a metal-rich peak near $[\mathrm{M/H}]=-0.5$. We also observe a global offset between the WFM and NFM distributions. The origin of this offset is currently unclear. We speculate it could be a result of the different noise characteristics of the NFM and WFM data. The AO correction in the NFM degrades towards the blue cut-off of the MUSE wavelength range (where many metal lines are located), resulting in enhanced noise in the blue part of the extracted spectra. Consequently, the \texttt{Spexxy} fits of the NFM spectra may be more tightly constrained by the red spectral features than those of the WFM spectra, and the strength of the strong calcium triplet lines tends to be underestimated in the \texttt{GLib} spectra.

To characterise the offset, we fit each distribution with a double-Gaussian model. The fitted NFM peaks lie above their WFM counterparts by $0.125$~dex for the metal-poor component and $0.222$~dex for the metal-rich component. We adopt the WFM metallicity scale as our reference. To align the metal-poor peaks, we subtract $0.125$~dex from the metallicity of every NFM star.

After this correction, a double-Gaussian fit to the combined sample yields metal-poor and metal-rich peaks at $[\mathrm{M/H}]=-1.58$ and $-0.53$, respectively. We assign each star probabilities $p_1$ and $p_2$ of belonging to the metal-poor and metal-rich components, respectively, with $p_1+p_2=1$. We classify stars with $p_1>0.5$ as metal-poor and those with $p_1\leq0.5$ as metal-rich. Before the shift, the fitted $p_1=0.5$ boundary lies at $[\mathrm{M/H}]=-1.09$ for the WFM subsample and $-0.87$ for the NFM subsample. After the shift, the fitted boundary for the combined sample lies at $[\mathrm{M/H}]=-1.03$ (right panel of Fig.~\ref{fig:me_hist_corrected}).
 
Both metallicity peaks have standard deviations of $\sigma_{\rm [M/H]}\approx0.2$, which are higher than the upper limit of 0.1 we imposed on our ${\rm [M/H]}$ measurements. This could imply that the peaks have intrinsic metallicity spreads. However, the formal uncertainties of quantities measured by \textsc{Spexxy} from MUSE spectra are known to be underestimated \citep{2016A&A...588A.148H}. \citet{2023ApJ...958....8N} used repeat observations of $\omega$~Cen to empirically calibrate corrections to the metallicity uncertainties returned by \texttt{Spexxy}, and found that underestimated measurement errors contributed to the observed scatter. Without multi-epoch data of the cluster, it is difficult to assess whether the widths of the two metallicity peaks visible in Fig.~\ref{fig:me_hist_corrected} could be explained by measurement uncertainties. We defer this analysis to a future publication.

We highlight that, so far, we have only considered the cool stars in the MUSE sample, as the hot subsample defined in Sec.~\ref{sec:muse} could not be used for reliable metallicity measurements. In light of the bimodality among the cool stars depicted in Fig.~\ref{fig:me_hist_corrected}, we divide the MUSE sample into three stellar groups for the subsequent analyses, namely a \emph{metal-poor} population composed of cool stars within the metallicity peak at $[\mathrm{M/H}]=-1.6$, a \emph{metal-rich} population composed of cool stars within the metallicity peak at $[\mathrm{M/H}]=-0.5$, and a \emph{young} population containing the stars bluer than F606W-F814W$=0.6$.

\subsection{Cluster Centre and Number Density Profile}
\label{sec:morphology}

\begin{figure}
    \centering
	\includegraphics[width=1.\columnwidth]{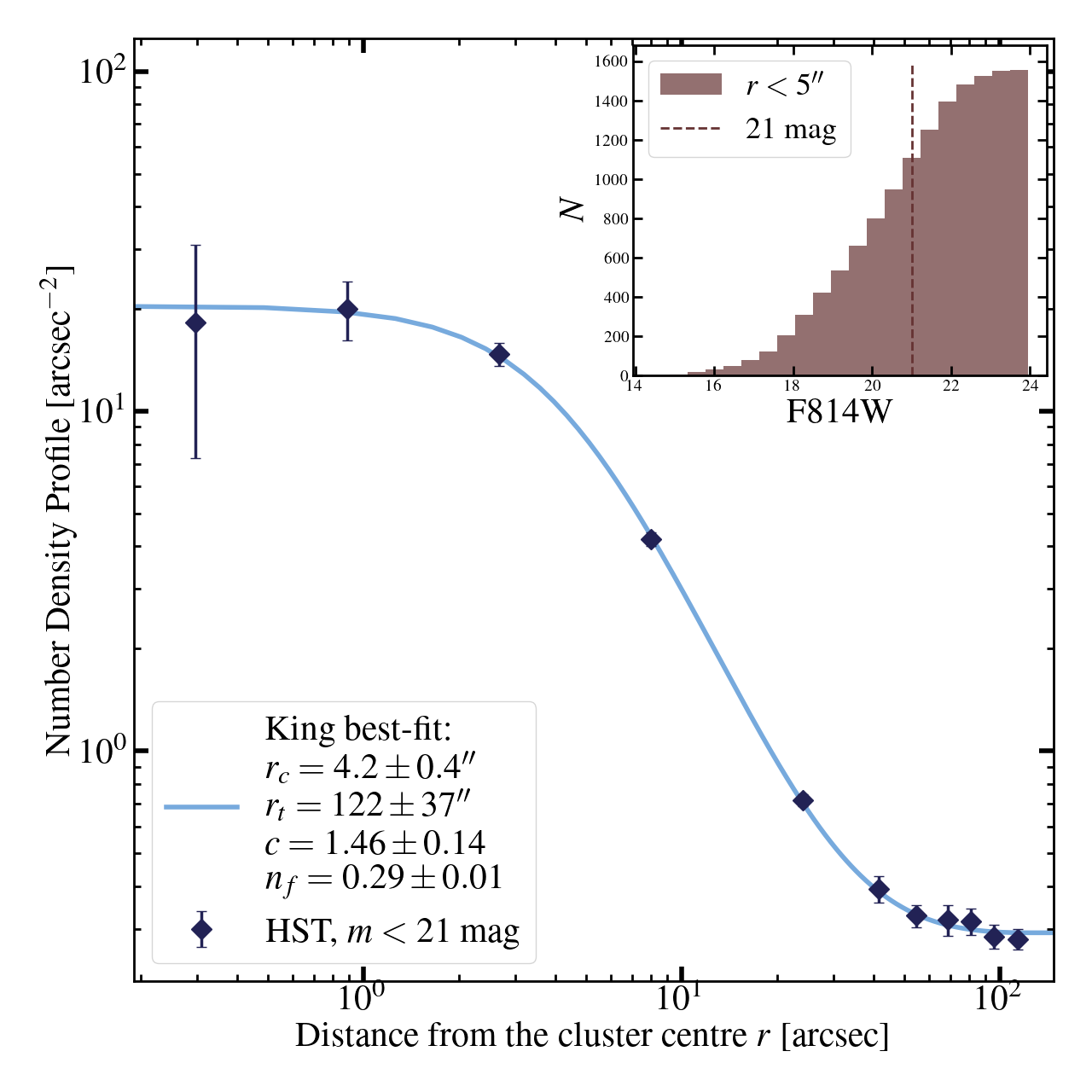}
    \caption{The measured NGC~1916 number density profile is presented by dark blue diamonds, with uncertainties computed from Poisson noise. The King model fit to the data, including a background component, is shown by the light blue solid line. The parameters of the best-fit profile are provided in the legend. A cumulative histogram of the HST number counts as a function of F814W magnitude for stars within 5~arcsec of the cluster centre is shown in the upper right panel. The adopted magnitude cut is indicated by a dashed vertical line.}
    \label{fig:ndp}
\end{figure}

We used the HST photometry to infer the centre of NGC~1916, following the method of \citet{2010ApJ...710.1032A} as detailed in Appendix~\ref{sec:centre}. We find central values of $\alpha=05^{\rm h}18^{\rm m}37.61^{\rm s}$ ($79.65669 \pm 0.00003$~deg) and $\delta=-69^\circ24^\prime25.56^{\prime\prime}$ ($-69.40710 \pm 0.00002$~deg), which will be used throughout the following analyses.

In the next step, we measured the number density profile (NDP) of NGC~1916, as detailed in Appendix~\ref{sec:ndp}. Only stars brighter than F814W$=21$ were included in the calculation. This number represents a compromise between having sufficient star counts within the radial bins and avoiding incompleteness in the HST photometry at even fainter magnitudes. As illustrated in Fig.~\ref{fig:ndp}, the result is reasonably well matched by a \citet{1962AJ.....67..471K} profile with a concentration of $c=1.46$ and a core (or King) radius of $r_{\rm c}=4.2"$, corresponding to 1.00~pc at an assumed distance to the LMC centre of 50~kpc.

The King-profile fit further allows us to estimate the density of field stars across the cluster. We find an absolute number surface density of stars brighter than our threshold (of F814W$=21$) of $\Sigma_{f\rm }=0.29~{\rm arcsec}^{-2}$, corresponding to 1.41\% of the central number surface density of the King profile.

\subsection{Membership and Internal Kinematics}
\label{sec:kinematics}

\begin{figure*}
    \centering
	\includegraphics[width=\textwidth]{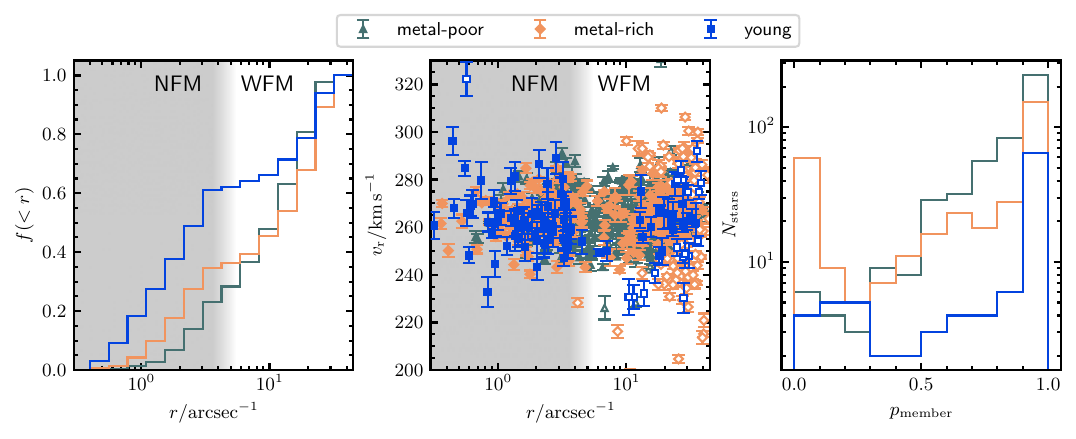}
    \caption{Membership probability determination for the MUSE data. \textit{Left:} Raw cumulative number counts of the stars in the three subgroups defined in Sec.~\ref{sec:me} as a function of distance to the cluster centre, prior to any membership selection. \textit{Centre:} Line-of-sight velocities as a function of distance from the cluster centre for the same subgroups. Stars are plotted as filled or open symbols depending on whether their cluster membership probabilities are above or below 50\%. We note the presence of a young star which is located close to the cluster centre, but classified as a field star due to its high velocity. We discuss this peculiar object further in Sec.~\ref{sec:imbh}. \textit{Right:} Histograms of the membership probabilities for the three subgroups. All three subgroups are dominated by cluster members, whereas the field stars are mainly metal rich. In the left and central panels, the radial coverage of the NFM data is indicated by grey shadings.}
    \label{fig:membership}
\end{figure*}

We aim to use the line-of-sight velocities derived from the MUSE data to perform a kinematic selection of likely members of NGC~1916. To this end, we used a maximum-likelihood method similar to the one described in Sec.~3 of \citet{2013MNRAS.436.2598W}, where each star is assigned a membership probability based on its velocity and projected distance from the cluster centre. We used the King-profile fit described in the previous section and Appendix~\ref{sec:ndp} to assign membership priors to all stars in the MUSE sample via
$$ P_{\rm memb}^{\rm prior}\sim \frac{\Sigma_{\rm c}}{\Sigma_{\rm c}+\Sigma_{\rm f}},$$
where $\Sigma_{\rm c}$ is the surface number density of the cluster at a given radius (see Appendix~\ref{sec:ndp}) and $\Sigma_{\rm c}+\Sigma_{\rm f}$ is the total surface number density of stars at the same radius. In the case of NGC~1916, we found $\Sigma_{\rm f}$ to be 1.41\% of the central cluster surface number density. The stellar kinematics are then fitted with a double-Gaussian distribution: one Gaussian represents the cluster component, which is typically narrower and more populated, while the second Gaussian represents the field population, which usually shows a higher dispersion. While the velocity dispersion of the field stars is assumed to be constant, the dispersion amongst the cluster stars is modelled to decrease with projected distance to the centre according to a King profile. We used \textsc{Limepy} \citep{2015MNRAS.454..576G} to convert the surface density profile depicted in Fig.~\ref{fig:ndp} into a velocity dispersion profile. To scale the velocity dispersion profile, we allowed the cluster mass to vary during the analysis, while the structural parameters were held fixed.

To optimise the parameters of the model, we used a maximum-likelihood approach based on Markov-chain Monte Carlo using \textsc{emcee} \citep{2013PASP..125..306F}, fitting for the cluster mass, the systemic velocities of the cluster and field populations, and the velocity dispersion of the latter. For consistency with the analysis of the surface number density profile, $\Sigma_f$ was held fixed. An overview of the model parameters, their fitted values and previous measurements is provided in Table~\ref{tab:fit}, while we include a corner plot of the fitted parameters in Appendix~\ref{sec:corner}. For convenience, we also provide the central velocity dispersion of the best-fit King model, which follows from the cluster mass. The kinematic parameters of the LMC field are in good agreement with the measurements of \citet{2002AJ....124.2639V}. The line-of-sight velocity of NGC~1916 obtained in our analysis, $264.5 \pm 0.4~\kms$, differs by $\sim7~\kms$ from the value of $271.5 \pm 1.5~\kms$ reported by \citet{2019MNRAS.482.1275U} based on integrated-light spectroscopy. The origin of this offset is unclear, but might be related to a minor offset in the wavelength calibrations. We note, however, that \citet{1997A&A...324..505D} obtained $267.6 \pm 0.8~\kms$ analysing integrated-light spectra from ESO/CASPEC, which yields a better agreement with our result. It is also worth noting that we do not find a significant difference between the systemic velocities of the cluster and the field population, as one might have expected if the cluster was located behind or in front of the bulk of field stars. 

For the central velocity dispersion $\sigma_{\rm c,\,LOS}$, we obtain a value of $10.1^{+0.3}_{-0.5}~\kms$. This is slightly above the value of $8.0\pm0.6~\kms$ reported by \citet{1997A&A...324..505D} from integrated-light spectroscopy.

\begin{table}
    \centering
    \begin{tabular}{l|c|c}
    Parameter [units] & Best fit & Literature \\
    \hline
    Cluster velocity $v_{\rm c,\,LOS}$ [km/s]& $264.5 \pm 0.4$ & $271.5 \pm 1.5^a$ \\
    Cluster mass $M_{\rm cl}$ [$10^5~{\rm M_\odot}$] & $4.2^{+0.4}_{-0.3}$ & $7.1\pm1.4^b$ \\
    Cluster central dispersion $\sigma_{\rm c}$ [km/s] & $10.1^{+0.3}_{-0.5}$ & $8.0\pm0.6^c$ \\
    Field velocity $v_{\rm f\,LOS}$ [km/s] & $263.8 \pm 1.6$ & $262.2\pm3.4^d$ \\
    Field dispersion $\sigma_{\rm f}$ [km/s] & $21.5 \pm 1.1$ & $20.2 \pm 0.5^d$\\
    \end{tabular}
    \caption{MCMC best-fit results for the dynamical model of NGC~1916 and the surrounding field. $a$ -- \citet{2019MNRAS.482.1275U}, 
    $b$ -- the Galactic Globular Cluster Database \citep{2018MNRAS.478.1520B}, 
    $c$ -- \citet{1997A&A...324..505D}, $d$ -- \citet{2002AJ....124.2639V}. }
    \label{tab:fit}
\end{table}

\begin{figure}
    \centering
	\includegraphics[width=\columnwidth]{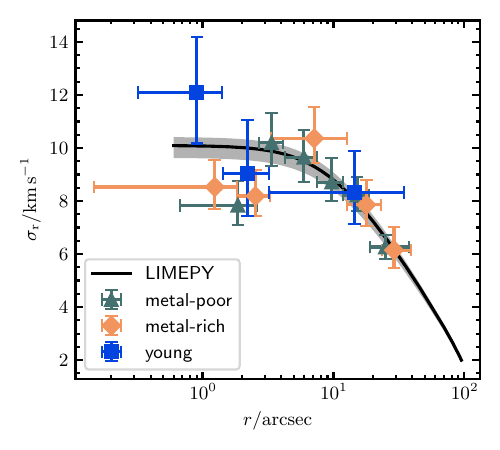}
    \caption{Kinematic profiles. The solid line shows the best-fit velocity dispersion profile of the overall cluster population obtained with \texttt{LIMEPY} for a King profile, with a shaded band indicating the $1\sigma$ confidence interval. Radially binned velocity dispersion profiles are overplotted for the three stellar populations identified within NGC~1916, using the same colour-coding as in previous figures. All populations follow the overall profile, suggesting that field contamination is minimal. In all cases, a membership probability cut of $p > 0.5$ has been applied (cf. Sec.~\ref{sec:memb}).}
    \label{fig:dispersion}
\end{figure}

We then selected stars with a posterior membership probability greater than $0.5$ as likely cluster members, resulting in 131 of the 905 stars in our sample being classified as likely field stars and discarded from further analyses. The results of the cluster member selection as a function of stellar group are summarized in Fig.~\ref{fig:membership}, where the left panel shows the cumulative radial distributions of the three groups prior to any membership cuts. It is remarkable that the metal-poor RGB population is the \emph{least} centrally concentrated of the three, which would not be expected if the other two were dominated by LMC field interlopers. This impression is corroborated in the middle panel of Fig.~\ref{fig:membership}, where the velocities of the stars in the three groups are plotted as a function of distance to the cluster centre, with open symbols highlighting likely field stars. At large radii, a substantial number of young stars and especially metal-rich RGB stars are flagged as field stars. Such an excess of predominantly metal-rich field stars is consistent with expectations for the LMC, given the recent star formation activity of the galaxy \citep{2009AJ....138.1243H, 2021MNRAS.508..245M, 2022MNRAS.513L..40M}, and seems in agreement with the shape of the field CMD shown in Fig.~\ref{fig:cmd1}. However, even after accounting for this contamination, substantial intra-cluster populations of young stars and metal-rich RGB stars remain. This is also visible in the right panel of Fig.~\ref{fig:membership}, showing histograms of the posterior membership probabilities derived for all groups. While most of the stars with low membership probabilities are found in the metal-rich RGB group, the majority of the stars in this group do have membership probabilities close to one. The same is true for the young population. The metal-poor RGB population shows the lowest fraction of field stars, suggesting that there is no significant population of metal-poor RGB stars in the field surrounding NGC~1916.

Following the removal of field stars, we investigated the velocity dispersion profiles of the three stellar groups. To obtain the dispersion profiles, we created radial bins, each covering at least a radial range of $d\log(r/1^{\prime\prime})>0.2$ and a number of 50 stars in the cases of the RGB populations and 25 stars in the case of the young populations. We then used a similar MCMC approach to determine the dispersion, but this time we assumed the dispersion to be constant within each bin and fixed the mean velocity to the one listed in Table~\ref{tab:fit}. The result is shown in Fig.~\ref{fig:dispersion}. We see that all three populations follow the global King model within the uncertainties, further solidifying the evidence that the cluster hosts different stellar populations. We note that both RGB populations show a central dip in their dispersion profiles. A potential reason for this dip could be velocity biases due to crowding, given the extreme stellar surface densities right around the cluster centre. However, it is interesting to note that the young population does not show such a dip and instead appears to reveal a central dispersion cusp. Hence, it also seems possible that the young population dominates the stellar populations near the very centre of the cluster where the dispersion is expected to be highest. To solidify this trend, however, additional observations would be needed, in order to improve the accuracies of the velocity dispersion measurements and to investigate the impact of binary stars (which are expected to be more frequent among the young stars).

\subsection{Overdensities of Metal-Rich and Young Stars}
\label{sec:memb}

\begin{figure*}
    \centering
	\includegraphics[width=\textwidth]{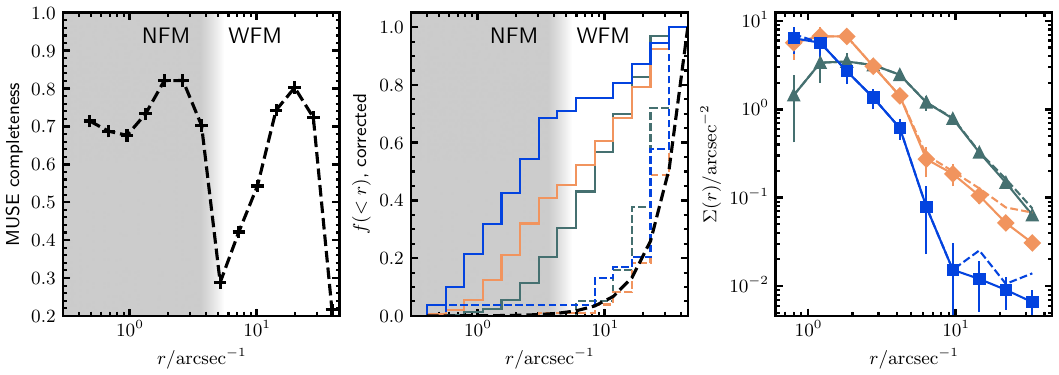}
    \caption{Radial profiles of the different subpopulations of NGC~1916 identified in the MUSE data. \textit{Left:} Completeness of the MUSE data above F814W=19 as a function of distance to the cluster centre. The coverage of the NFM data is indicated using grey shading. \textit{Centre:} Cumulative number counts as a function of cluster centre distance, following correction for incompleteness. Solid histograms are used for likely member stars ($p_{member} > 0.5$) and dashed histograms for likely field stars. The black dashed line shows the expected distribution of field stars when assuming a constant field-star density. \textit{Right:} Number density profiles following correction for incompleteness.}
    \label{fig:completeness}
\end{figure*}

In this section, we show clear evidence that, in addition to the well known metal-poor population in NGC~1916, a metal-rich population of red stars as well as young blue stars are both present in the cluster, and in fact are more concentrated than the metal-poor population.

We calculate the radial completeness of the MUSE data relative to the HST catalogue for all stars with $F814W < 19$. This is shown in the left-hand panel of Fig.~\ref{fig:completeness}. Three regimes are clearly seen; in the central arcsecond the completeness is $\sim$70\% due to the high crowding level of the data\footnote{We note that right at these radii, the HST data also become incomplete; this likely accounts for the apparent rise in completeness within the central arcsecond.}. The completeness then rises quickly to larger radii in the NFM data as the density drops. However, the edge of the NFM field of view (shaded grey vertical fade) results in a loss of completeness as the lower-spatial-resolution WFM data remains crowded at these radii. Then at radii $>10\arcsec$, the completeness improves again until the edge of the WFM data is reached.

Using the radial completeness curve, we can correct the raw number counts shown in the left panel of Fig.~\ref{fig:membership} by dividing the number count in each bin by the corresponding value of the radial completeness curve (obtained via linear interpolation). We do so separately for the members and non-members, and show the result in the middle panel of Fig.~\ref{fig:completeness}. Selecting just non-members ($p_{member} < 0.5$) results in number density profiles that rise significantly starting at a radius $>$10\arcsec{}. The consistency of these profiles with the expected constant background profiles suggests our membership probabilities are successfully identifying non-member stars in all three populations.

The cumulative profiles of member stars (with $p_{member} > 0.5$; solid lines in the middle panel of Fig.~\ref{fig:completeness}) clearly show that the young colour-selected blue stars show the most concentrated profile, while the metal-poor stars show the least concentrated. The right panel shows the number density of member stars showing similar behaviour. These make it clear that the density of young and metal-rich stars eclipses the metal-poor population at the centre, although the downward turn in the central number density of metal-poor stars is likely due to incompleteness.

As shown in the right panel of Fig.~\ref{fig:membership}, the majority of non-member stars are metal-rich. Thus it is critical to understand whether we are seeing a true overdensity of metal-rich stars in the cluster, or whether these might just be foreground stars. As an independent test of our kinematic membership probabilities, we calculate the total number of field stars expected in the NFM data based on the HST field star density at bright magnitudes. Using the field star density of $207 \pm 14$ per $1 \times 1$~arcmin$^2$ with F814W$<$19 estimated from HST data (Sec.~\ref{sec:hst}), we expect just 3.2 field stars in the NFM field of view (without accounting for completeness, which would lower the number further). The observed number of metal-rich stars in the NFM data is $113$ --- a factor of $\sim$35 above the expected field level. This makes it clear that the overdensity of metal-rich stars is statistically highly significant.

Turning to the young stellar component, we can identify an HST-based sample of stars from the colour-magnitude diagram. As seen in Fig.~\ref{fig:cmd1}, the cluster CMD shows a clear overdensity of blue stars at F814W$\sim$20 that are consistent with being a blue/extreme horizontal branch population from the metal-poor population similar to those seen in other LMC clusters \citep{1998MNRAS.300..665O}. However, there are also a number of brighter blue stars in the cluster CMD that are likely younger, massive stars. We examine the location of these young massive stars across the HST field by using the same colour and magnitude cuts used to select the young stars in the MUSE data: F814W$<$19 and F606W-F814W$<$0.6, thus excluding the possible horizontal branch population. These stars show a clear compact clump within the central parts of the NFM field of view, with $59$ bright blue stars within $3''$ (0.7~pc) of the cluster centre. At larger radii, the bright blue stars are relatively uniformly distributed across the full HST field of view, with a density in the background regions of $38$ stars per square arcminute, translating into an expectation of just $0.3$ bright blue stars in the central 3 arcseconds of the cluster. Thus, as with the metal-rich component highlighted above, these stars' association with the cluster is highly significant. 

We also note that visually, the MUSE NFM cube (Fig.~\ref{fig:footprint}, right panel) shows a number of bright blue stars in the central 3\arcsec{} of the cluster; $14$ of these stars show clear H$\alpha$ emission, visible as green-coloured stars in the NFM image. These bright emission-line stars have F814W magnitudes between 18.7 and 16.7 and have spectra consistent with being Be stars, including relatively broad H$\alpha$ emission (FWHM$\sim$10\AA) and modest Helium absorption at $4921{\text \AA}$ \citep[e.g.,][]{2020A&A...634A..51B}. We also note the H$\alpha$ emission in the WFM data (Fig.~\ref{fig:footprint}, middle panel), which shows a clear cavity centred on NGC~1916. This is likely also a signature of hot stars in the cluster with the bubble being created by the impact of their stellar winds on the surrounding medium. 

This measurement firmly establishes that the cluster contains at least three distinct populations of stars: a metal-rich and metal-poor population of cooler stars and a population of bright, blue, hot stars. We discuss more details of these three coincident stellar populations within NGC~1916 in Section~\ref{sec:stellarpops} below.

\subsection{Systemic Proper Motion}
\label{sec:pm}

\begin{figure}
    \centering
	\includegraphics[width=0.9\columnwidth]{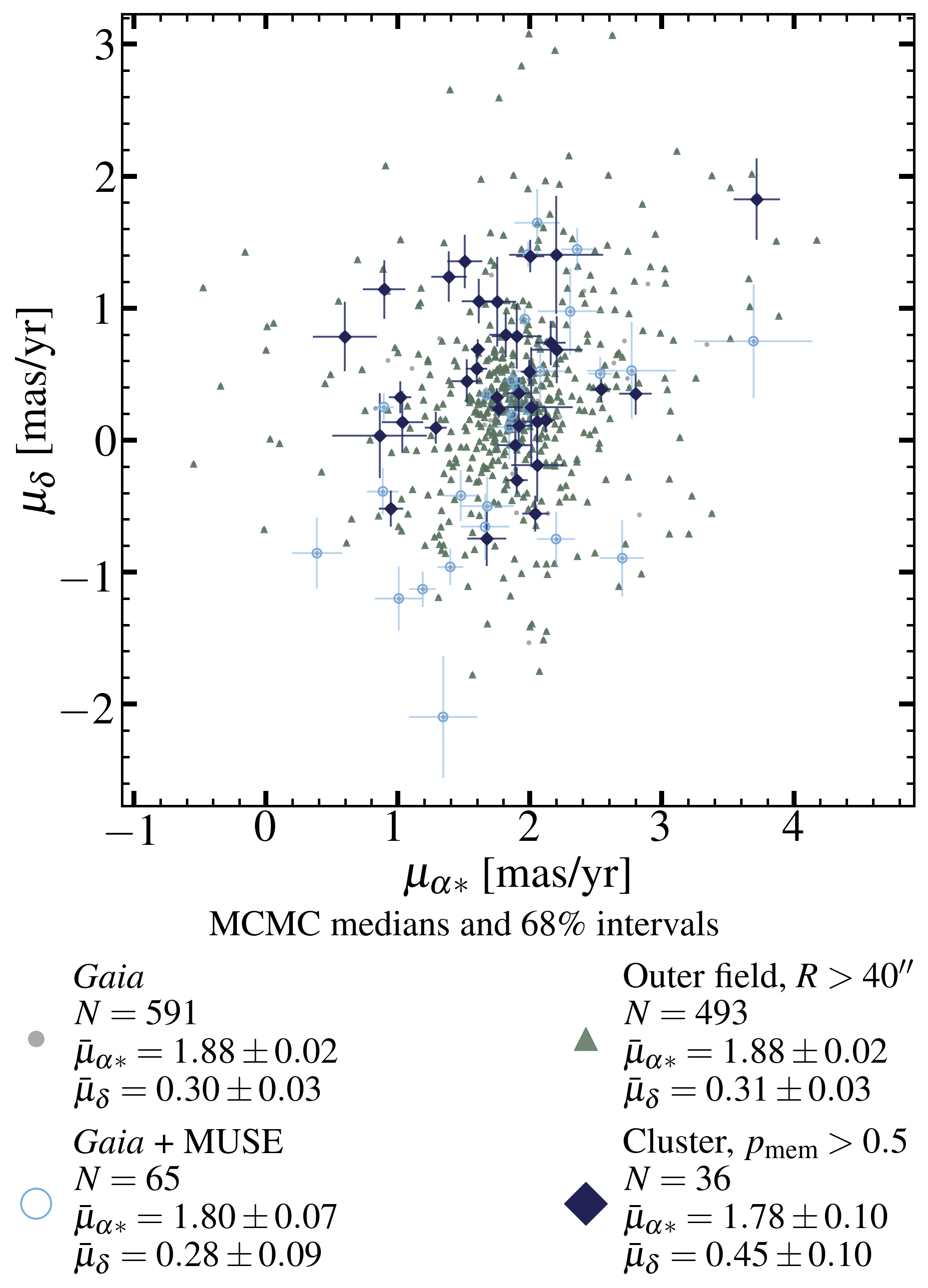}
    \caption{Proper motions of 591 \textit{Gaia} stars in the HST field of NGC~1916. Grey points show the full sample, and green triangles the 493 outer-field stars at projected radii $R>40\arcsec$. Open light blue circles mark the 65 stars with MUSE counterparts. Filled dark blue diamonds identify the 36 likely cluster members with $p_{member}>0.5$ based on MUSE kinematics (Sec.~\ref{sec:kinematics}). Error bars show the individual \textit{Gaia} proper-motion uncertainties for the MUSE counterparts only. The legend gives the mean motion of each sample and its 68\% credible interval in each coordinate. The cluster and outer-field mean motions show a slight offset, but are consistent in both coordinates, within the $2\sigma$ credible interval.}
    \label{fig:gaia}
\end{figure}

Making use of Gaia DR3 proper motions (cf. Sec.~\ref{sec:gaia}), we examined the velocity distribution of stars within the HST field of view. Due to source crowding, the individual proper motion uncertainties are substantial, ranging from $\sim$$12$ to nearly $100$~km~s$^{-1}$, which greatly exceeds the internal velocity dispersion measured from the MUSE line-of-sight velocities. Nevertheless, the data are sufficient to compare the cluster bulk motion with that of the surrounding LMC field. 

We used MCMC to estimate the mean proper motions of three samples, accounting for the individual \textit{Gaia} measurement uncertainties. Of the 905 stars in our MUSE sample, 65 have \textit{Gaia} counterparts and 36 of these have cluster membership probabilities $p_{member}>0.5$. All 36 likely members were observed in WFM. In the order $(\mu_{\alpha}\cos\delta,\,\mu_{\delta})$, we find $(1.88\pm0.02,\,0.30\pm0.03)$ for the full sample of 591 \textit{Gaia} stars, $(1.79\pm0.07,\,0.29\pm0.09)$ for the 65 stars with MUSE counterparts, and $(1.77\pm0.10,\,0.48\pm0.10)$ for the 36 likely cluster members (Fig.~\ref{fig:gaia}). All proper motions are in $\mathrm{mas\,yr^{-1}}$. We report posterior medians, with uncertainties given by the half-widths of the central 68\% credible intervals. The results are shown in Fig.~\ref{fig:gaia}. Stars cross-matched with members from the MUSE catalogue appear to exhibit a lower dispersion relative to the field population, as expected for bound cluster members.

For comparison, we selected 493 \textit{Gaia} stars at projected radii $R>40\arcsec$ from the cluster centre. This outer-field sample is disjoint from the 36 likely cluster members and has a mean proper motion of $(1.88\pm0.02, 0.31\pm0.03)~\mathrm{mas\,yr^{-1}}$. Subtracting the cluster mean from the outer-field mean gives an offset of $+0.12~\mathrm{mas\,yr^{-1}}$ in $\mu_{\alpha}\cos\delta$ (95\% credible interval $[-0.09,+0.33]~\mathrm{mas\,yr^{-1}}$) and $-0.17~\mathrm{mas\,yr^{-1}}$ in $\mu_{\delta}$ (95\% credible interval $[-0.39,+0.04]~\mathrm{mas\,yr^{-1}}$). The posterior probability of a positive offset in right ascension is 86.6\%. Both offsets are consistent with zero within their 95\% credible intervals. Also taking the agreement in the systemic line-of-sight velocities into account (see Table~\ref{tab:fit}), we do not find evidence that the cluster moves relative to the surrounding field population.

\section{Discussion}
\label{sec:discussion}

We have found above that at least three clearly distinct stellar populations are present in NGC~1916, with red metal-poor and metal-rich stars as well as young, blue stars, many of which have H$\alpha$ emission. The presence of young stars along with older stars with a bimodal metallicity distribution is unique amongst clusters in the LMC, and strongly suggests that NGC~1916 represents the NSC of the LMC. In this section, we first discuss the stellar populations in further detail, examining their distribution on the colour-magnitude diagram, providing rough age estimates, and examining their spatial distributions. 
We then derive a dynamical mass estimate for NGC~1916, and place it in the context of the NSC--host galaxy mass relation. Then we examine NGC~1916 in the context of other NSCs, and close by discussing the implications of the finding in the context of the LMC. 

\subsection{The Complex Stellar Populations of NGC~1916}
\label{sec:stellarpops}

\begin{figure*}
    \centering
 	\includegraphics[width=0.99\columnwidth]{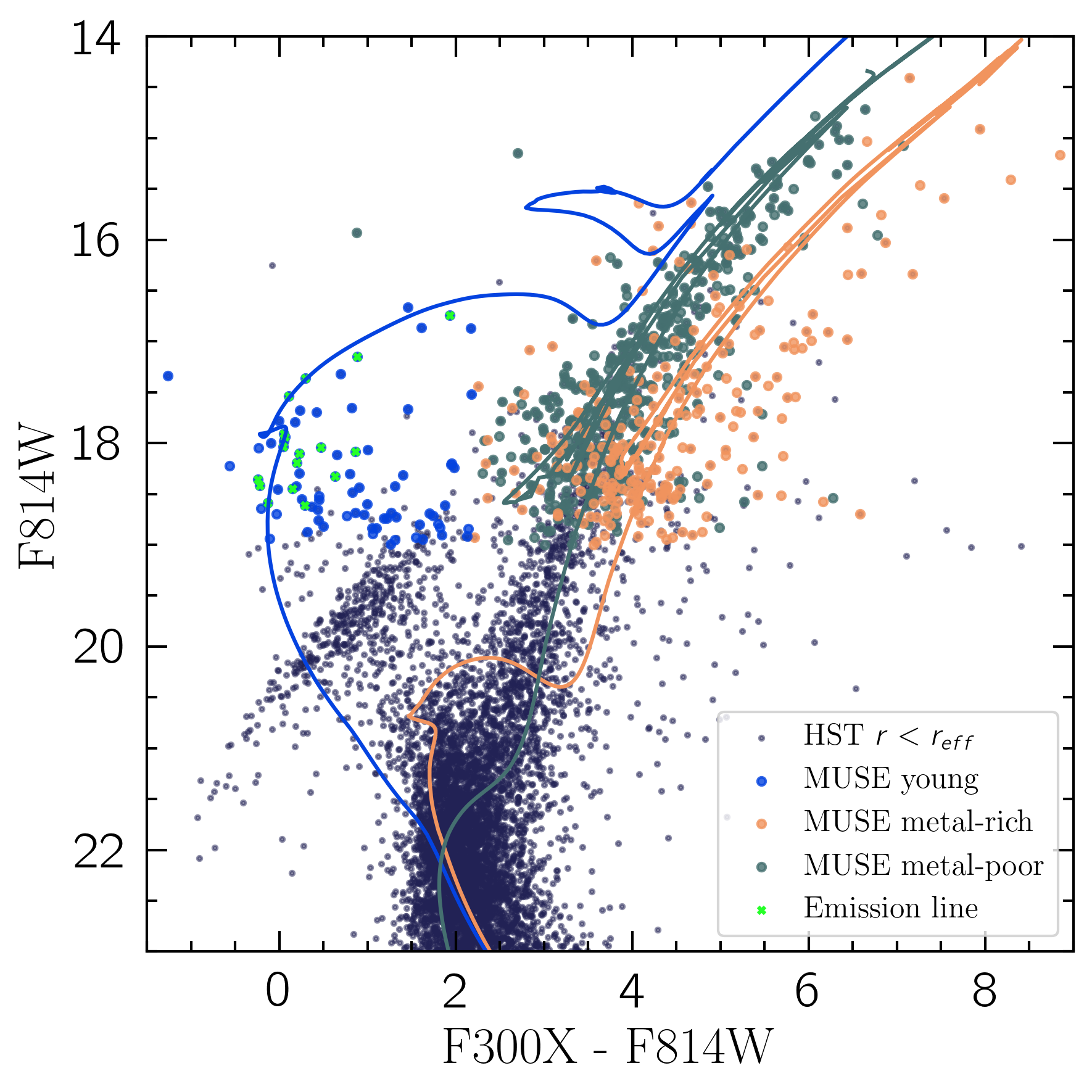}
     \includegraphics[width=0.99\columnwidth]{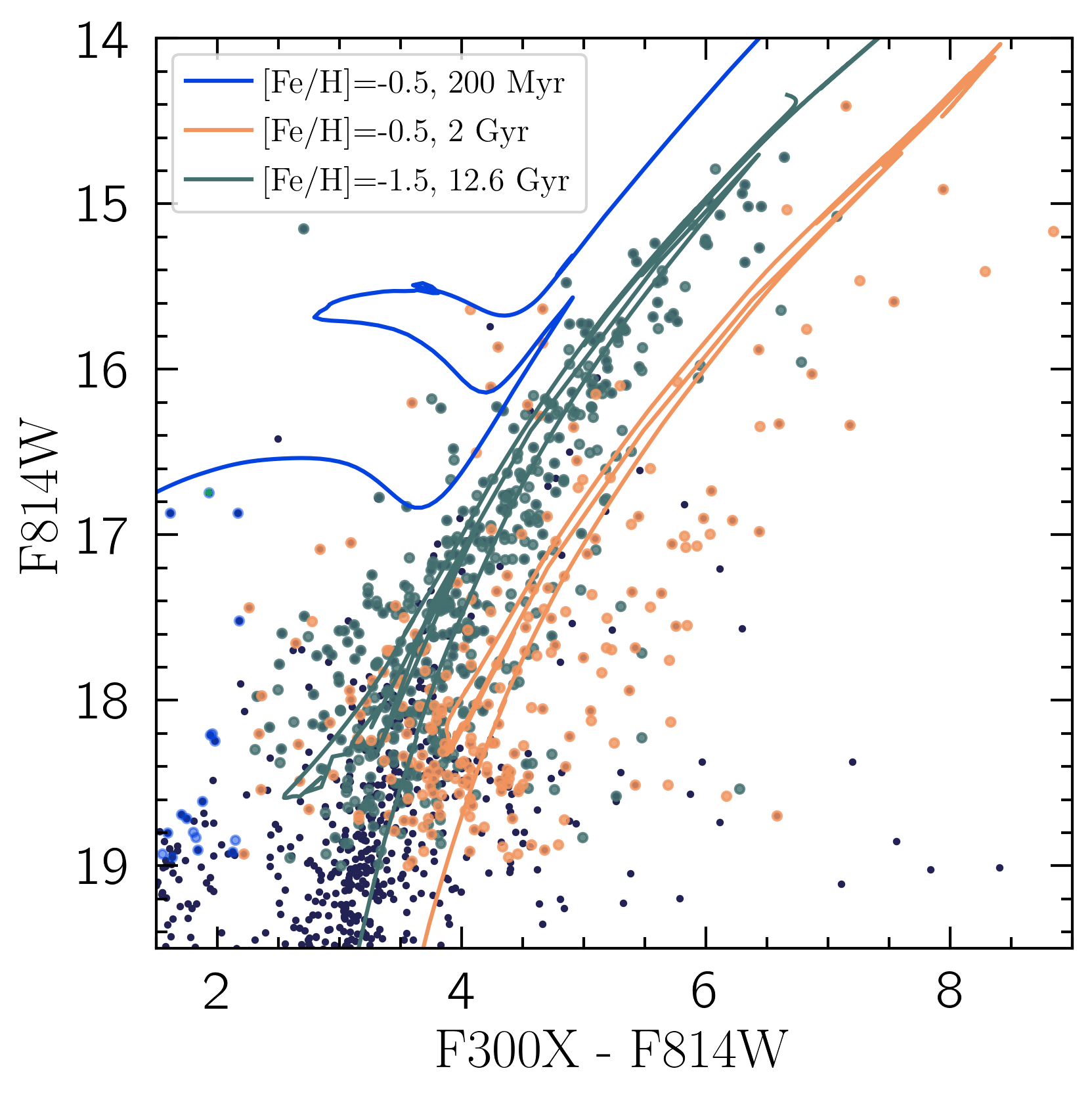}
\caption{The complex stellar populations of NGC~1916. Both panels show all HST stars within the half-light radius of $11.7''$. Overplotted are MUSE stars with $p_{member} > 0.5$. The left panel shows the full CMD down to the old main sequence turn-off, while the right panel focuses on the red giant branch of the cluster where spectroscopic measurements are available for a significant fraction of sources. Overplotted are MIST isochrones \citep{2016ApJS..222....8D,2016ApJ...823..102C} at metallicities of $-1.5$ and $-0.5$, roughly matching the peaks of the MUSE-derived metallicities, and with ages selected to match features within the CMD.}
\label{fig:cmd_wiso}
\end{figure*}

{\bf Stellar Ages:} In Fig.~\ref{fig:cmd_wiso} we show the F300X and F814W cluster colour-magnitude diagram as derived with HST within the half-light radius with the MUSE measurements highlighted by colour. The left panel shows the full CMD. At the bright blue end, two features are clearly seen: a diagonal streak of blue stars around F814W$\sim$20 which are consistent with being blue or extreme horizontal branch stars and then a smaller number of brighter stars following the upper main sequence of the 200~Myr isochrone shown (note that we assume an $A_V = 0.58$ for all isochrones).\footnote{We note that some stars included in our sample of young stars are consistent with being horizontal branch stars. However, their relative number ($\sim$10/99) is too small to impact the conclusions of our work.} The brightest blue stars imply at least some stars that have likely ages $\sim100$~Myr, with the brightest Be star also suggesting a similar age \citep{Bastian2017}. The large number of Be stars evident from Fig.~\ref{fig:cmd_wiso} may appear surprising, however, young star clusters in the Magellanic Clouds typically reach Be-star fractions around 50\% near their turn-offs \citep{2020A&A...634A..51B,2025MNRAS.542.2768K}. These stars' metallicities are not reliably obtained from the MUSE spectra due to a lack of metal absorption lines, but given their youth, it is likely their metallicity is similar to the present-day gas phase metallicity in the LMC of [O/H]$\sim-0.3$ \citep[e.g.][]{Lah2024}. Any blue stars of populations with an age of $\sim$100s of Myr should be accompanied by a population of red helium burning stars which may be present just to the left of the RGB \citep{Kerber2007}; a handful of metal-rich stars around 16th magnitude are consistent with this population, with additional stars seen around 18th magnitude, which suggest a somewhat older age of $\sim$600~Myr.  

The main sequence turn-off is quite thick, with the oldest turn-off stars at around F814W$\sim$22 consistent with the old metal-poor isochrone, but with a clear overdensity of stars to the upper left of this up to F814W$\sim$21. On the red giant branch (right panel), the metal-rich and metal-poor populations clearly separate based on colour in this filter combination (while they overlap significantly in F606W-F814W). The same old (12.6~Gyr) metal-poor isochrone used for our initial guesses in spectral fitting (cf. Sec.~\ref{sec:muse}) provides a reasonable fit to the metal-poor population, while a $\sim$2~Gyr isochrone matches the position of the red clump at F814W$\sim$18.5, and the brighter main-sequence turn-off overdensity at F814W of 21. Older isochrones may help explain the RGB stars redwards of this isochrone but large populations of stars with ages $\gtrsim$5~Gyr are ruled out as they would produce a fainter and redder main-sequence turnoff and subgiant branch than are observed.  
Overall, it appears the metal-rich stars have an extended star formation history extending from the young ages ($\sim$100~Myr) to account for the bright blue stars, and to at least above 1~Gyr to create the metal-rich red clump and RGB stars seen in the MUSE data.

{\bf Spatial Distribution:} we also compare the spatial distributions of the different populations. Based on the King profile fit to number counts in Section~\ref{sec:morphology} the overall cluster has a half-light radius of $11.7''$ (2.84~pc); although incompleteness at the centre may make this a bit of an overestimate. We use the completeness corrected cumulative distributions of member stars in the middle panel of Fig.~\ref{fig:completeness} to get the relative concentrations of the three populations of stars. The blue, young population is the most concentrated with a half-light radius of MUSE members of just 3.3\arcsec{} (0.80~pc). The metal-rich RGB and red clump stars are also significantly more concentrated than the cluster as a whole with a half-number radius of 7.6\arcsec{} (1.84~pc). The metal-poor stars have a half-light radius of 12.3\arcsec{} (2.98~pc), just slightly larger than the half-light radius from the King profile fit to the number density profile. The total number of member stars is dominated by the metal-poor component, this is likely also the dominant component by mass, in agreement with the existing metal-poor metallicity inferred from integrated light spectra by \citet{Colucci2012}.

Our inference of the stellar populations is significantly complicated by the combined impacts of metallicity spreads, age spreads, and differential reddening towards the cluster. Ultimately, all three factors must be considered to fully explain the complex RGB morphology of NGC~1916. We will more quantitatively explore the stellar populations of the cluster and their morphology using a new HST program on NGC~1916 (PID: 18136, PI: Seth) in a future paper. These data will provide proper motion measurements, thus leading to more refined membership criteria, and F438W photometry enabling construction of reddening-free colours.

\subsection{Cluster Mass Estimation}
\label{sec:mass}

In Sec.~\ref{sec:kinematics}, we used a \textsc{Limepy} model to infer a dynamical cluster mass of $4.2\times10^{5}~M_{\rm \odot}$. We can compare this value to the virial mass estimate of the cluster. Following \citet{2002AJ....124.2615L}, we calculate the virial mass as
$$ M_{\rm vir} = \frac{7.5\,r_{\rm h}\,\sigma_\infty^2}{G},$$
where $\sigma_\infty=8.3^{+0.3}_{-0.3}~\kms$ is the cluster-wide line-of-sight velocity dispersion and $r_h$ is the King model’s three-dimensional half-mass radius. Thus, a total mass of the cluster is $M_{\rm tot}=(4.30^{+0.34}_{-0.29})\times10^5~\Msun$, where the quoted uncertainties correspond to the 68\% interval from the velocity fit in Sec.~\ref{sec:kinematics}. 

Our dynamical mass measurements are lower than previous estimates. \citet{2003MNRAS.338...85M} determined a photometric mass of $(5$--$6)\times10^5~\Msun$, while the online Galactic Globular Cluster Database of \citet{2018MNRAS.478.1520B} gives a total mass of $7.1\times10^5~\Msun$ for NGC~1916 based on $N$-body modelling, about 1.7 times our value. This places NGC~1916 at the low end of the empirical \mbox{NSC--host} galaxy mass relation. For the LMC stellar mass of $M_\star \approx 2.7 \times 10^9~\Msun$, the observed scaling relations predict an NSC mass of $\sim\!3$--$7 \times 10^6~\Msun$ \citep{2016MNRAS.457.2122G, 2020A&ARv..28....4N, 2023MNRAS.520.4664H}, approximately one order of magnitude above our estimate. The offset is, however, within 2$\sigma$ of the $\sim$0.5~dex intrinsic scatter of these relations, and is consistent with the systematically lower NSC masses observed in late-type galaxies at fixed host mass \citep{2016MNRAS.457.2122G}.

\subsection{Stellar Population Comparison with M~54 and $\omega$~Cen and other Nuclear Star Clusters}
\label{sec:comparison}

The stellar populations of NGC~1916 match the expectations and observations of other NSCs quite well. The presence of a combination of old metal-poor stars and younger metal-rich stars is seen in other NSCs, with a transition around a galaxy stellar mass of $\sim$10$^9$~M$_\odot$ (similar to the LMC) from metal-poor stars dominating the NSCs to metal-rich stars dominating; galaxies in the intermediate mass range similar to the LMC in particular show significant mass fractions of both populations \citep{2020A&ARv..28....4N, 2021A&A...650A.137F, 2022A&A...658A.172F,2025ApJ...979...85L}. The old metal-poor populations that dominate the NSCs of lower mass galaxies are thought to be due to the inspiral of globular clusters due to dynamical friction, while the younger metal-rich stars are thought to be due to recent formation from gas. Like in NGC~1916, the youngest stars in the Milky Way NSC \citep{FeldmeierKrause2015}, the Andromeda NSC \citep{Bender2005, Lauer2012}, and other NSCs \citep{Carson2015} are centrally concentrated on similar physical scales ($<$1~pc) to what we see in NGC~1916.

We can also compare the stellar populations to two other stripped NSCs that we have resolved stellar data for: M~54 and $\omega$~Cen. The latter shows a large metallicity spread ($-2.5 < \mathrm{[Fe/H]} < -0.6$; \citealt{2010ApJ...722.1373J, 2017ApJ...844..164B}) but a notably smaller age spread of only $\sim3$~Gyr \citep{2024ApJ...977...14C}. In contrast, M~54, the NSC of the Sagittarius dwarf galaxy, hosts two distinct stellar populations with a significant age spread: an old metal-poor component ($\sim12$~Gyr, $\mathrm{[Fe/H]} \approx -1.4$) and a young metal-rich component ($\sim2$~Gyr, $\mathrm{[Fe/H]} \approx 0$; \citealt{2019ApJ...886...57A}) \footnote{A third population identified by \citet{2019ApJ...886...57A} with intermediate age and metallicity ($\sim4$~Gyr, $\mathrm{[Fe/H]} \approx -0.3$) likely represents the Sgr field stars.}. Kinematic analysis revealed that the old population shows low rotation, consistent with ex-situ formation, while the young population displays significant rotation, suggesting in-situ formation from accreted gas \citep{2020ApJ...892...20A}. A comparison of NGC~1916 with these two clusters is presented in Table~\ref{tab:comparison}. 

The following might be considered a possible evolutionary path for NGC~1916. Its metallicity spread of $\sim1$\,dex is comparable to that of M~54, and the presence of clearly separated metal-poor and metal-rich populations suggests a substantial age spread between the two. The fact that the metal-rich population has a metallicity comparable to the young field stars of the LMC suggests it formed recently. At the same time, some of these metal-rich stars lie on the RGB, implying ages of at least $1$\,Gyr. The LMC experiences ongoing interactions with the SMC and the Milky Way that recently boosted its star formation \citep{2009AJ....138.1243H, 2022MNRAS.513L..40M}, providing support for a hybrid scenario, similar to M~54, where the metal-poor population resembles a classical globular cluster and the metal-rich population formed when fresh gas was funnelled to the cluster centre.


Future work on the stellar populations with new HST data from program 18136 will enable more detailed constraints on the formation history of the cluster.

\begin{table}
    \centering
    \begin{tabular}{l|c|c|c}
    Property & M~54 & $\omega$~Cen & NGC~1916 \\
    \hline
    Mass [$\Msun$]$^*$ & $1.6 \times 10^{6}$ & $4 \times 10^{6}$ & $7.1\times 10^{5}$ \\
    $\mathrm{[Fe/H]}$ range [dex] & $-1.4$ to $0$ & $-2.5$ to $-0.6$ & $-1.5$ to $-0.5$ \\
    Age spread [Gyr] & $\sim10$ & $\sim3$ & $\sim10$ \\
    Host galaxy & Sgr dSph$^a$ & $\omega$ Dwarf$^b$ & LMC \\
    \end{tabular}
    \caption[Table 2]{Comparison of NGC~1916 with other NSC candidates. $^*$Masses are taken from the Galactic Globular Cluster Database \citep{2018MNRAS.478.1520B} to ensure consistency, as all values are derived using the same N-body fitting methodology. $a$ -- \citet{1994Natur.370..194I}, $b$ -- \citet{2026ApJ..1008..158S}.} 
    \label{tab:comparison}
\end{table}

The lack of direct age measurements for NGC~1916 prevents us from fully characterising its star formation history. However, the metallicity bimodality alone strongly suggests an extended formation timescale, as self-enrichment from a single stellar population cannot produce the observed $\sim1$~dex spread. Future deep photometry reaching the main-sequence turn-off region would enable age determinations for the distinct populations and allow a more detailed comparison with the age--metallicity relations established for M~54 and $\omega$~Cen. 

A similar breadth of metallicity is characteristic of NSCs in host galaxies of comparable mass to the LMC. Integrated light spectroscopy of NSCs in nearby spiral and early-type galaxies with stellar masses $2-8 \times 10^9 \Msun$ reveals prolonged star formation histories spanning more than 1.5~dex in metallicity, from metal-poor old populations formed more than 10~Gyr ago to enriched populations at later epochs \citep{2018MNRAS.480.1973K}. At the lower end of this mass range, \citet{2022A&A...667A.101F} find diverse star formation histories among NSCs in late-type dwarf galaxies with stellar masses $\sim 10^9 \Msun$, where contributions from both GC inspiral and in~situ star formation produce comparably broad metallicity distributions. The metallicity bimodality of NGC~1916, hosted by a galaxy of stellar mass $2.5 \times 10^9 \Msun$, is therefore consistent with the range of stellar population properties observed in NSC analogues at similar host masses, further supporting an extended, multi-episode formation history.

\subsection{Implications for the LMC and its Massive Black Hole}
\label{sec:imbh}

The ability of NSCs to form stars {\em in situ} over time is the result of their position at the centres of galaxies. This may imply that NGC~1916 was at the dynamical centre of the LMC as recently as $\sim$100~Myr ago and that it may still mark the dynamical centre of the LMC. A wide range of dynamical centre positions have been inferred for the LMC (see Fig.~\ref{fig:centres}, Table~\ref{tab:lmc_centres}); only those from \citet{2014ApJ...781..121V} show consistency with NGC~1916 within their 1$\sigma$ error bars. Alternatively, NGC~1916 may not be at the present-day dynamical centre of the galaxy, perhaps related to the ongoing dynamical interaction with the Milky Way. A recent paper by \citet{Chisolm2026} suggests that dynamical interactions with the Milky Way can induce up to $\sim$1$^\circ$ (900~pc) offsets for the LMC's central black hole; given the compact distribution of mass in NGC~1916, a similar offset could be present, and explain the possible present-day offset of NGC~1916 from the LMC's dynamical centre.

While offset black holes are expected in dwarf galaxies due to galaxy-galaxy interactions \citep[e.g.][]{Bellovary2021}, the known nearby central black holes in galaxies with masses similar to that of the LMC have been found at the centres of their NSCs \citep[e.g.][]{Nguyen2019,Davis2020}. Furthermore, dynamical interactions may cause both the NSC and central black hole to slosh around together. This suggests that if the LMC hosts a massive black hole, it may be in NGC~1916.  

The most direct evidence for a massive black hole in the LMC comes from hypervelocity stars \citep{Erkal2019, 2025ApJ...982..188H}. These hypervelocity stars are young stars with masses above 2~M$_\odot$, thus the presence of a centrally concentrated young massive population of stars in NGC~1916 is consistent with the possible generation of hypervelocity stars in the cluster. \citet{Lucchini2025} trace back the orbits of the hypervelocity stars to infer the possible location of the LMC black hole -- NGC~1916 lies at the edge of their 2$\sigma$ positional inference for the location of the black hole. \citet{2025ApJ...982..188H} also use the number and velocity of the stars to infer the black hole mass in the LMC; they infer a mass of $\sim$6$\times$10$^5$~M$_\odot$. Such a massive black hole could clearly not be located in NGC~1916, as it is larger than the entire dynamical mass inferred from the cluster (Sec.~\ref{sec:mass}). However, a lower-mass black hole could still be hiding in the cluster; the dispersion of the young stars appears to increase at the centre despite the impacts of crowding, which typically reduces the velocity dispersion from individual stars due to blending with spectra lying near the systemic velocity of the cluster. With our new HST observations from program 18136 (PI: Seth), we will both refine the likely position of the LMC black hole using proper motion measurements of two hypervelocity stars, and get deeper kinematic measurements at the centre of NGC~1916 from proper motions. We will dynamically model NGC~1916 with these to get a strong constraint on any possible black hole at the centre of the cluster. 

The strongest indication of a black hole would be high-velocity stars near the centre of the cluster that feel the black hole's potential \citep{2024Natur.631..285H}. One possible candidate of a fast star is visible in the middle panel of Fig.~\ref{fig:membership}, namely the young star at a projected distance of $<1~{\rm arcsec}$ and with a high velocity of approximately $+60~\kms$ relative to the cluster mean. However, this star shows very strong H$\alpha$ emission (it is the brightest green dot in the right panel of Fig.~\ref{fig:footprint} and the brightest grey data point along the blue isochrone of Fig.~\ref{fig:cmd_wiso}) and additional line emission on top of H$\beta$ and the Paschen lines. Hence, its velocity measurement is uncertain. By fitting the H$\alpha$ emission with a Gaussian profile using \textsc{lmfit} \citep{newville_2025_16175987}, we obtain a velocity of $274\pm3~\kms$, which is much more in line with the systemic velocity of NGC~1916. We conclude that the star is likely a Be star and a cluster member, yet does not suggest the presence of a massive black hole. We note that owing to its velocity offset, the star was considered a field star during our analyses and did not artificially inflate the velocity dispersion profile of the young stars shown in Fig.~\ref{fig:dispersion}. That the young stars have the highest dispersion and are the most centrally concentrated of the three populations suggests that any high-velocity stars are most likely to be young, in agreement with observations at the Galactic Centre.

\section{Conclusions}
\label{sec:conclusions}

We have presented a combined photometric and spectroscopic analysis of NGC~1916, the most massive star cluster in the central region of the Large Magellanic Cloud. 


The MUSE integral-field spectroscopy, obtained in both wide-field and narrow-field modes, provides spectra of 905 stars. We combine a King profile fit of the number counts of stars in the cluster with the MUSE kinematics of the stars to determine cluster membership. The cluster line-of-sight velocity is centred at $v_{\mathrm{LOS}} \approx 265$~km\,s$^{-1}$ with a central velocity dispersion of $\sigma_{\mathrm{c}} \approx 10$~km\,s$^{-1}$.  
Along the red giant branch, we find a bimodal metallicity distribution for member stars. A dominant old ($\sim$12~Gyr) metal-poor population peaks at $\mathrm{[M/H]} \approx -1.6$~dex, while a secondary metal-rich population peaks at $\mathrm{[M/H]} \approx -0.5$~dex; the metal-rich population shows a clear red clump indicating stellar ages of at least 1~Gyr are present. In addition to these older stars, nearly a hundred young massive stars are found within the cluster with ages as young as $\sim$100~Myr. Many of these show emission lines indicating they are Be stars, similar to those seen in other young clusters in the LMC. While the field stars are also metal-rich, the density of the metal-rich RGB stars in the central NFM data is a factor of $\sim$35 higher than the expected background of field stars, and the young blue star overdensity is also highly significant. The young blue stars are the most concentrated population, with the half-light radius of the young stars being just 0.8~pc; the metal-rich RGB stars show a somewhat broader distribution, with a half-light radius of 1.8~pc. The metal-poor red giant branch stars are the least concentrated, with a half-light radius of $\sim$3~pc, similar to the cluster as a whole. We estimate the mass of the cluster using our King profile fit and stellar dispersion measurements and find a mass of 4.2$\times$10$^5$~M$_\odot$. The kinematics of the different populations are mostly consistent with each other, with the youngest stars having the highest central velocity dispersion, consistent with their high concentration.

The presence of these distinct stellar populations in NGC~1916 is similar to what is seen in other NSCs. In addition, the central concentration of young stars is regularly seen in other NSCs, similar to the Milky Way, Andromeda and other nearby nuclei in galaxies with similar mass to the LMC. Our ability to detect individual stars in NGC~1916 facilitates direct comparison to tidally stripped NSCs from dwarf galaxies in the Milky Way, M~54 and $\omega$~Cen. Like M~54, NGC~1916 appears to host both an old, metal-poor population -- likely formed through ex-situ processes such as globular cluster inspiral -- and a younger, metal-rich component potentially formed in situ, driven by gas accretion towards the LMC centre. That might be a result of a boosted star formation episode caused by SMC or MW interactions. This formation scenario is consistent with observed populations in NSCs within the LMC mass range around $10^{9}$~$\Msun$ where both channels appear to contribute \citep{2020A&ARv..28....4N, 2021A&A...650A.137F, 2022A&A...658A.172F}. The mass of the cluster lies $\sim$1 order of magnitude below the expected NSC mass for a galaxy of the LMC's mass, but is within the scatter seen in other galaxies. Overall, the evidence strongly supports NGC~1916 as the NSC of the LMC.  

The identification of young massive stars concentrated in the centre of the cluster may also help explain the presence of B-type hypervelocity stars that appear to have been ejected from the LMC \citep{Erkal2019, 2025ApJ...982..188H}. The cluster's position is consistent, within 2$\sigma$, with the backwards integrated orbits of these stars \citep{2025ApJ...979...85L}, but the black hole mass inferred from the hypervelocity stars by \citet{2025ApJ...982..188H} is clearly inconsistent with the cluster velocity dispersions and dynamical mass. 

Data from an ongoing HST programme (18136; PI Seth) will provide additional photometric data and proper motions for the stars in NGC~1916. These will be used to explore the formation history and constrain the presence of a massive black hole in NGC~1916.


\section*{Acknowledgements}
EB acknowledges support through an STFC studentship.
SK gratefully acknowledges funding from UKRI through a Future Leaders Fellowship (grant no. MR/Y034147/1).
NB acknowledges the grants PID2024-155720NB-I00 and PID2024-159201NB-C21 funded by MCIN/AEI/10.13039/501100011033 as well as the Basque Government’s BERC and the Severo Ochoa Centre of Excellence Programmes (CEX2024-001491-S, funded by MICIU/AEI/10.13039/501100011033). 
JB is supported by an NWO Veni fellowship (VI.Veni.242.199).
We thank Peter Weilbacher for insightful discussions during the preparation of this manuscript.

\section*{Data Availability}
The raw HST and MUSE data are available in the respective archives. The HST photometric catalogue and the MUSE spectra, along with the results of the spectral analysis and the cluster membership determination, will be made available on Zenodo upon acceptance of the paper.

\bibliographystyle{mnras}
\bibliography{bibliography}

\appendix
\section{LMC Centres}
\label{sec:lmccentres}
Here we present an overview of all measured LMC centres obtained with different approaches and tracers. Table~\ref{tab:lmc_centres} compiles independent kinematic and dynamical centre measurements of the LMC from the literature. Early H\,{\sc i} studies placed the gas kinematic centre near $\alpha_0 \approx 78$--$79\degr$, significantly west of later stellar-based determinations. Stellar kinematic centres span a range of $\sim 3\deg$ in right ascension and $\sim 1\deg$ in declination, reflecting both methodological differences and the dependence on the tracer population used. In particular, \citet{2020MNRAS.492..782W} demonstrated that old (carbon stars and RGB stars) and young main-sequence stars each yield distinct centres offset by up to $\sim 1\deg$. Recent proper-motion and Jeans-modelling analyses based on \textit{Gaia} and VMC data have converged towards $\alpha_0 \approx 80.0$--$80.3\degr$, $\delta_0 \approx -69.3\degr$ \citep{2024AA...692A..40K, 2025AA...700A.279V}, though the scatter among all published values underscores that the LMC dynamical centre remains defined only to within $\sim 0.5\deg$. NGC~1916 occupies the same spatial region as expected for an NSC candidate of the LMC.

\begin{table*}
\caption{Independent kinematic/dynamical centre measurements of the LMC. All coordinates are J2000.}
\label{tab:lmc_centres}
\begin{tabular}{lccccll}
\hline
Reference & $\alpha_0$ [$^\circ$] & $\sigma_{\alpha}$ [$^\circ$] & $\delta_0$ [$^\circ$] & $\sigma_{\delta}$ [$^\circ$] & Tracers \\
\hline
\citet{1992AA...263...41L}$^a$       & 78.13 & --    & $-69.00$ & --        & H\,{\sc i} kinematics \\
\citet{1998ApJ...503..674K}$^a$      & 79.40 & --    & $-69.03$ & --        & H\,{\sc i} kinematics \\

\citet{2002AJ....124.2639V}      & 81.28 & $\pm0.24$ & $-69.78$ & $\pm0.08$ & Photometry \\
\citet{2002AJ....124.2639V}      & 81.91 & $\pm0.98$ & $-69.87$ & $\pm0.41$ & LOS of Carbon stars \\

\citet{2014ApJ...781..121V}      & 78.76 & $\pm0.52$ & $-69.19$ & $\pm0.25$ & PMs of 22 HST fields \\
\citet{2014ApJ...781..121V}      & 79.88 & $\pm0.83$ & $-69.59$ & $\pm0.25$ & PMs of HST fields, LOS of Old stars \\
\citet{2014ApJ...781..121V}      & 80.05 & $\pm0.34$ & $-69.30$ & $\pm0.12$ & PMs of HST fields and LOS of RSGs \\

\citet{2020MNRAS.492..782W}      & 80.90 & $\pm0.29$ & $-68.74$ & $\pm0.12$ & Gaia DR2 PMs of Carbon stars \\
\citet{2020MNRAS.492..782W}      & 81.23 & $\pm0.04$ & $-69.00$ & $\pm0.02$ & Gaia DR2 PMs of RGB stars \\
\citet{2020MNRAS.492..782W}      & 80.98 & $\pm0.07$ & $-69.69$ & $\pm0.02$ & Gaia DR2 PMs of Young MS stars \\

\citet{2021AA...649A...7G}$^a$   & 81.01 & --        & $-69.38$ & --        & Gaia eDR3 PMs of all populations \\
\citet{2022MNRAS.512.5423N}      & 79.95 & $\pm0.23$ & $-69.31$ & $\pm0.12$ & VMC PMs of intermediate/old stars \\
\citet{2024AA...692A..40K}       & 80.29 & $\pm0.04$ & $-69.25$ & $\pm0.02$ & Gaia 3D velocities of AGB, RSG \\
\citet{2025AA...700A.279V}       & 80.30 & $\pm0.02$ & $-69.27$ & $\pm0.02$ & VMC PMs of all populations \\
\hline
\multicolumn{6}{l}{$^a$ No formal errors reported for the centre.} \\
\end{tabular}
\end{table*}

\section{Centre of NGC~1916}
\label{sec:centre}

\begin{figure}
    \centering
    \includegraphics[width=0.9\columnwidth]{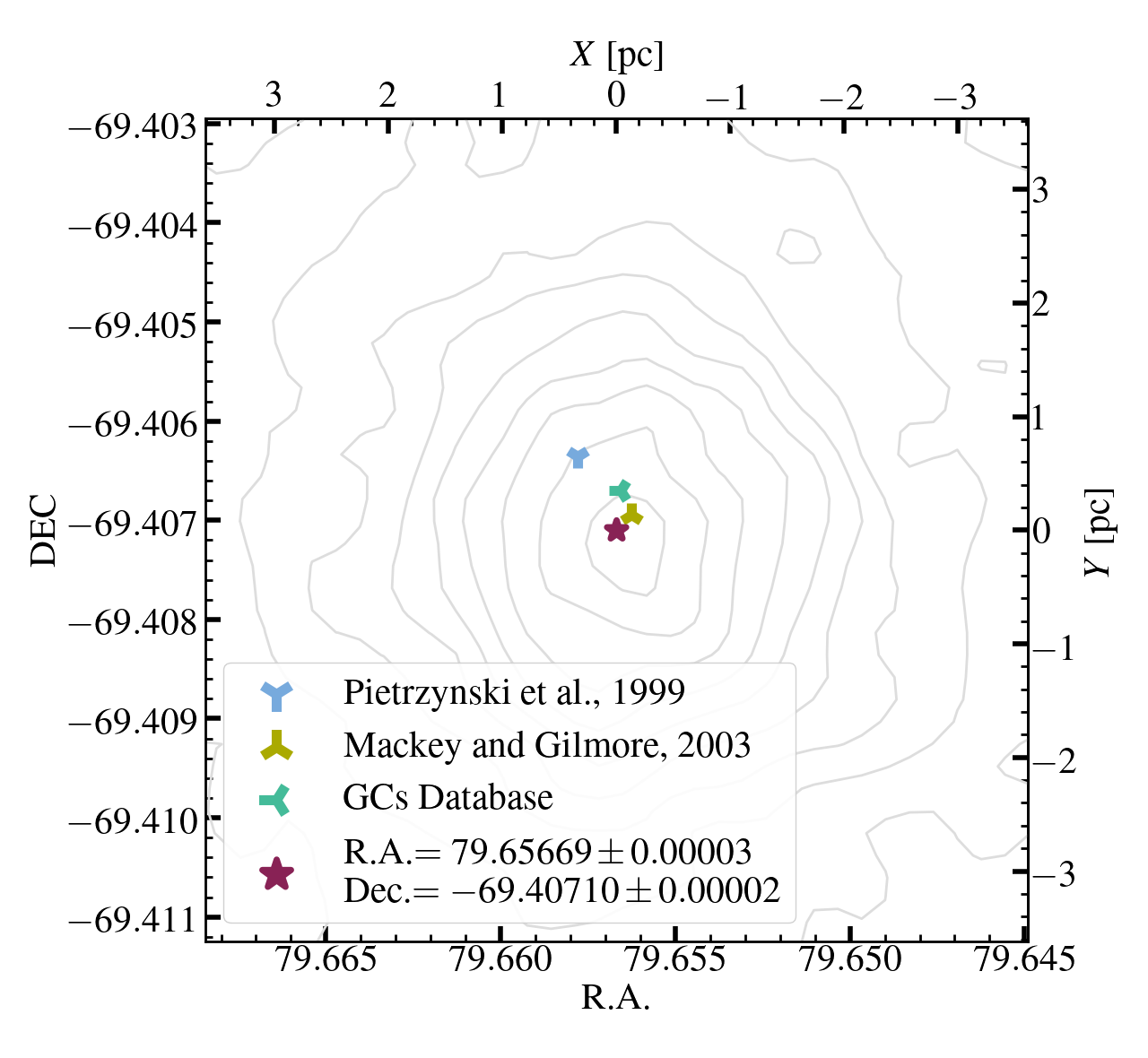}
    \caption{Best-fit central coordinates of NGC~1916 compared with previous measurements from \citet{1999AcA....49..521P, 2003MNRAS.338...85M} and the Galactic Globular Cluster Database \citep{2018MNRAS.478.1520B}. Grey contours show the projected stellar surface density of HST sources, constructed from a 2D histogram with smoothed 220 bins. Contour levels are equally spaced in surface density. The top and right axes give physical offsets in parsecs from our best-fit centre, assuming a distance to NGC~1916 of 50~kpc.}
    \label{fig:centralsearch}
\end{figure}

Being located in the central region of the LMC, NGC~1916 is expected to experience significant dust extinction. Moreover, the dust distribution across the cluster is irregular, making it difficult to determine its centre. While \citet{1998MNRAS.300..665O} presented WFPC2 photometry for NGC~1916, the cluster was excluded from the bulk of their analyses, including the centre determination, because of the strong differential reddening. \citet{1999AcA....49..521P} analysed a number of star clusters in the LMC. Using an interactive star count approach, they measured $\alpha=05^{\rm h}18^{\rm m}37.87^{\rm s}$ and $\delta=-69^\circ24^\prime22.9^{\prime\prime}$ for NGC~1916 centre.
\citet{2003MNRAS.338...85M} also studied a large sample of LMC globular clusters and measured their structural parameters, including the centres, using HST photometry. To locate the centres, they analysed the clusters' surface brightness distributions on a spatial grid and searched for the brightest point using a Monte Carlo approach. For NGC~1916, they obtained J2000 central coordinates of $\alpha=05^{\rm h}18^{\rm m}37.5^{\rm s}$ and $\delta=-69^\circ24^\prime25^{\prime\prime}$. 

Additionally, central coordinates $\alpha=05^{\rm h}18^{\rm m}37.572^{\rm s}$ and $\delta=-69^\circ24^\prime24.12^{\prime\prime}$ are listed at the Galactic Globular Cluster Database \citep{2018MNRAS.478.1520B}. We use them as our initial centre $\alpha_{0}$ and $\delta_0$.

To identify the centre of NGC~1916 based on our data, we apply the pie-slice method described in \citet{2010ApJ...710.1032A}. Since this method determines the centre as a location about which the stars are most symmetrically distributed, we have to remove any artificial asymmetry from the data. Thus, we restricted the analysis to stars brighter than $23$~mag in $\rm F814W$, as fainter stars showed spatial asymmetries that became particularly noticeable at $r>15$~arcsec from the initial cluster centre, likely due to an uneven reddening distribution.

The scatter for central coordinates from previous studies is $\Delta\alpha = 0.37^{\rm s} \approx 0.0015$~deg for $\alpha$ and $\Delta\delta = 2.1^{\prime\prime} \approx 0.0006$~deg for $\delta$, we set the initial range for trial centres as $[\alpha_{0}-\Delta\alpha; \alpha_{0}+\Delta\alpha]$, $[\delta_{0}-\Delta\delta; \delta_{0}+\Delta\delta]$ and set up a $100\times100$ grid that resulted in a sample of ten thousand coordinates and a resolution of $\sim 0.1^{\prime\prime}$. As the initial analysis resulted in a $\delta$ value close to the bottom limit, we increased the range for the centre search and used $[\alpha_{0}-\Delta\alpha; \alpha_{0}+\Delta\alpha]$, $[\delta_{0}-2*\Delta\delta; \delta_{0}-\Delta\delta]$ in the final analysis. 

For each trial centre, we perform the following:
\vspace{-\topsep}
\begin{enumerate}
    \item Measure the radial distance $r$ from the trial centre for each star at $r<15$'', and sort the sample by this value. Only stars within $15''$ of the trial centre were considered, to avoid the asymmetry due to extinction at larger radii.
    \item Divide stars into 8 equal slices and combine them into 4 pairs, each consisting of a slice and its opposite one.
    \item For each pair, identify the star closest to the centre in each slice and compute the difference $D$ in their radial distances $D = |r_1 - r_2|$.
    \item Repeat this procedure for $n = \min(N_1, N_2)$ stars, where $N_1$ and $N_2$ are the number of stars in each slice of the pair.
    \item Sum these differences for the pair of slices $D_{tot} = \Sigma_{n} D$.
    \item Combine pairs 1 and 3 (cardinal directions) and pairs 2 and 4 (semi-cardinal directions) separately.
    \item Sort trial centres by the total difference value $D_{tot}$.
    \item The trial centre with the smallest difference is adopted as the final centre.
    \item We define the uncertainty as the spatial separation between adjacent trial centres, i.e., our sampling resolution. The trial centre with the smallest total difference is adopted as the final centre, namely $\alpha=05^{\rm h}18^{\rm m}37.61^{\rm s}$ ($79.65669 \pm 0.00003$~deg) and $\delta=-69^\circ24^\prime25.56^{\prime\prime}$ ($-69.40710 \pm 0.00002$~deg).
\end{enumerate}

The determined centre and a comparison with previously measured centres are shown in Fig.~\ref{fig:centralsearch}. Overall, we find good agreement with the previous studies: our centre shows a slight offset of $\approx 0.8^{\prime\prime}$ compared to the one derived by \citet{2003MNRAS.338...85M}.

\section{Number Density Profile}
\label{sec:ndp}
\begin{figure}
    \centering
	\includegraphics[width=0.9\columnwidth]{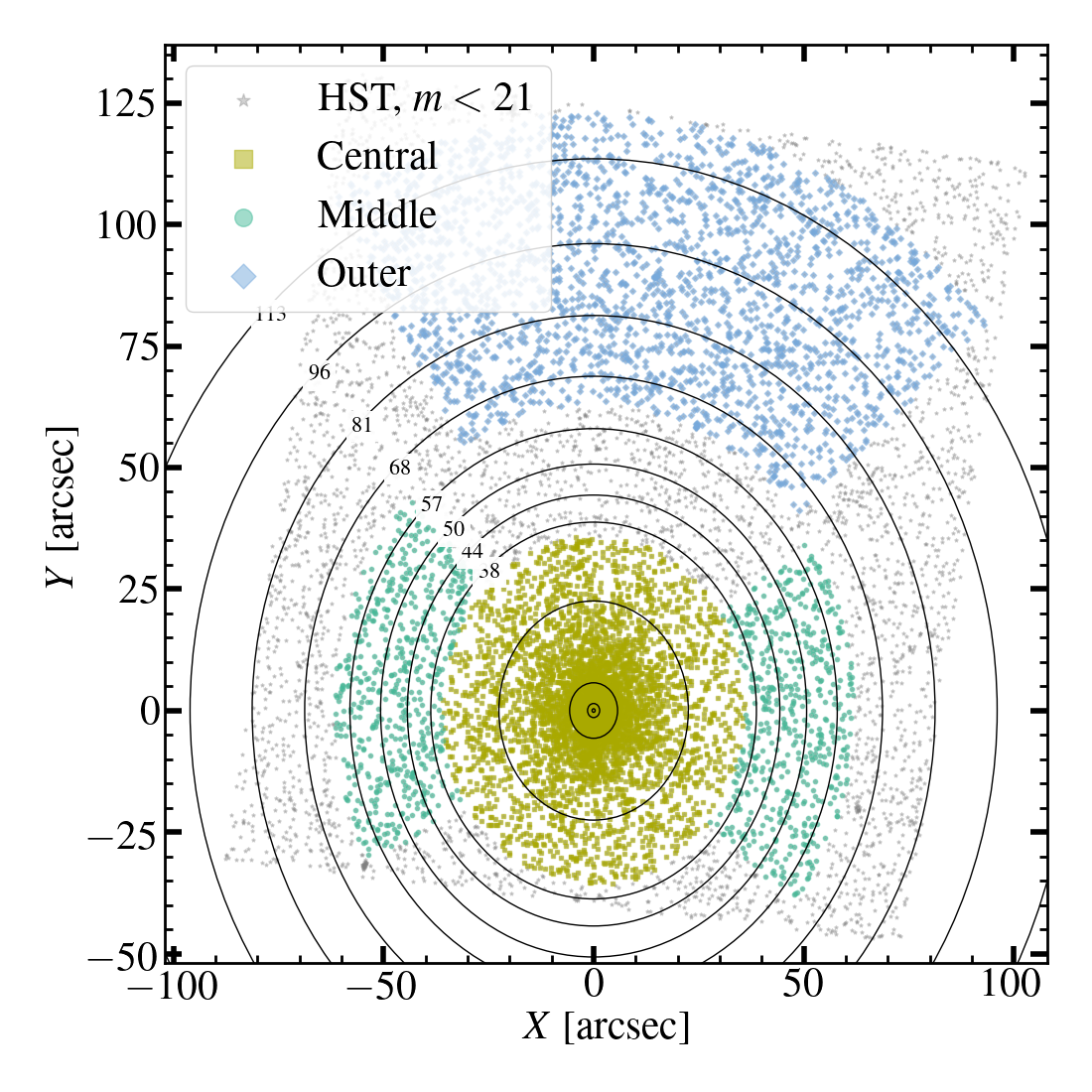}
    \caption{NGC~1916 on the HST footprint in grey, with highlighted areas that are used to measure the number density profile of the cluster. The central region, shown in olive, extends to $r = 36$~arcsec. The middle region, consisting of two sectors, spans radii from 36 to 62~arcsec, while the outer region covers $r = 62-123$~arcsec and is shown in blue. Black solid lines mark the radial bin centres used to construct the NDP shown in Fig.~\ref{fig:ndp}.}
    \label{fig:ndpsector}
\end{figure}

With the centre determined, we computed the number density profile (NDP) using stars with $m < 21$~mag. This magnitude cut was chosen based on the cumulative luminosity function within 5~arcsec of the cluster centre (inner panel of Fig.~\ref{fig:ndp}), which shows a turnover at $\sim 22$~mag indicating the onset of photometric incompleteness. 

The NDP was constructed by counting stars per radial bin on a logarithmic scale. Because the cluster is not centred within the HST pointing and there is a gap in the middle of the HST ACS field of view, we divided the field into three radial sectors with different levels of spatial coverage (Fig.~\ref{fig:ndpsector}) to avoid areas lacking photometric data. In the outer two sectors, star counts were multiplied by the inverse fraction of each annulus covered by the HST data. This approach allowed us to extend the NDP out to $\sim120$~arcsec and obtain an accurate background estimation. The constrained NDP is used for the membership analysis in Sec.~\ref{sec:kinematics}.

We fit the NDP with a \citet{1962AJ.....67..471K} model and include a background component. In this case, the functional form describing the number density as a function of projected distance to the cluster centre is
$$\Sigma(r) = \Sigma_{\rm f} + k \left[ \frac{1}{\sqrt{1 + (r/r_c)^2}} - \frac{1}{\sqrt{1 + (r_t/r_c)^2}} \right]^2,$$
where $k$ is the amplitude (scaling factor), $r_c$ is the core radius, $r_t$ is the tidal radius, and $\Sigma_{\rm f}$ is the background level.
The comparison between the measured profile and the best fit is shown in Fig.~\ref{fig:ndp}. The analysis yielded the following parameters:
\begin{enumerate}
    \item core radius $r_c = 4.2 \pm 0.4$~arcsec;
    \item tidal radius $r_t = 122 \pm 37$~arcsec;
    \item concentration value $c=1.46 \pm 0.14$;
    \item background level $\Sigma_{\rm f} = 0.29 \pm 0.01$ arcsec$^{-2}$.
\end{enumerate}
The tidal radius could only be loosely constrained by the King model fit, which is reflected in the large uncertainty quoted above. The corresponding effective radius for the best-fit model is 11.7 arcsec (2.84~pc). We also tested the \citet{1975AJ.....80..175W} profile, which yielded consistent results for the core radius, concentration parameter, and background level within the uncertainties, but similarly struggled to constrain the tidal radius. This is most likely due to insufficient spatial coverage, as the data do not extend far enough to constrain the outer boundary of the cluster.

\section{Metallicities with Ca~II}
\label{sec:metal}

\begin{figure}
        \centering
        \includegraphics[width=0.85\linewidth]{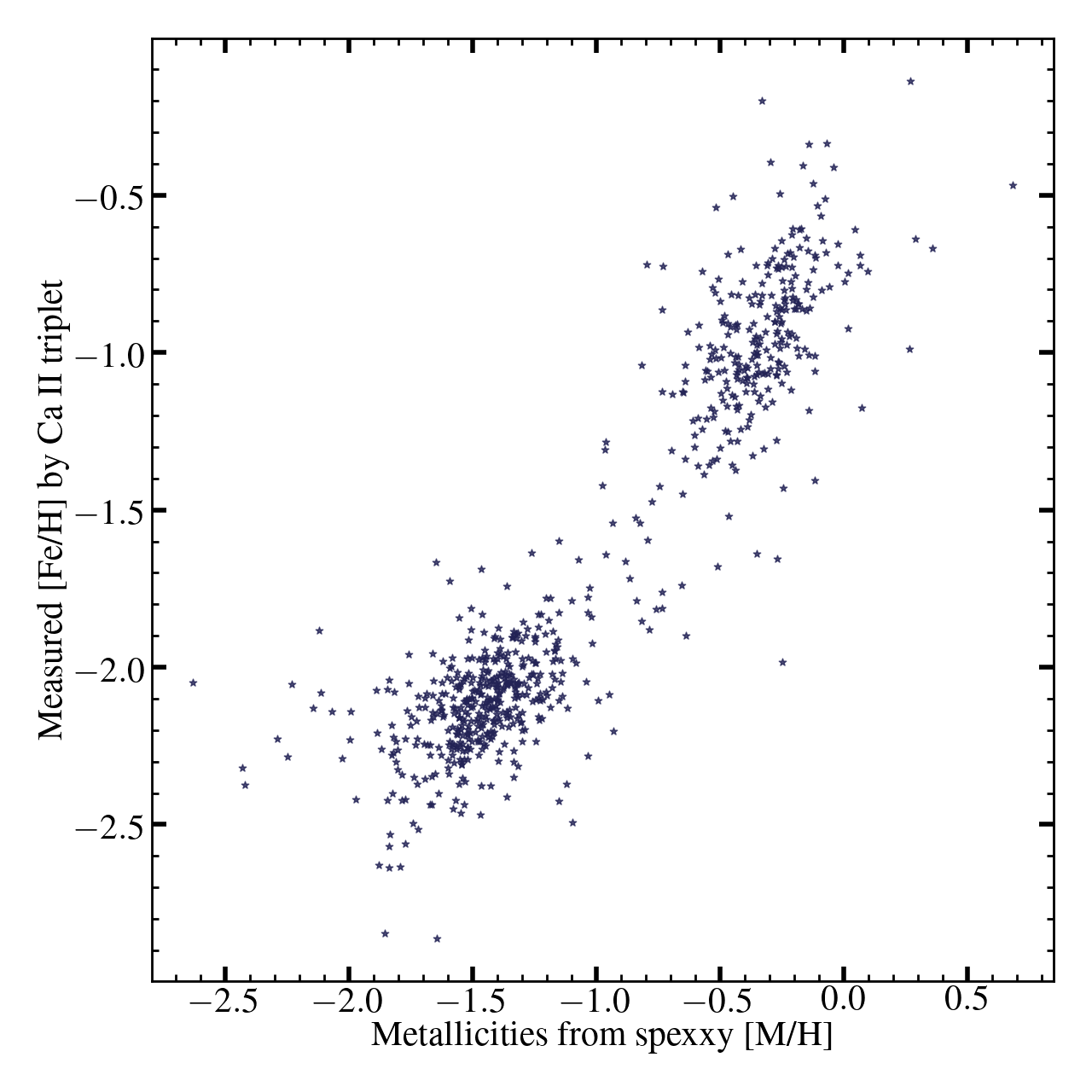}
        \caption{Comparison between the MUSE metallicities ${\rm[M/H]}$ derived from full spectrum fitting and ${\rm [Fe/H]}$ obtained from Ca~II triplet fitting.}
        \label{fig:mefe}
\end{figure}

To study the stellar populations of NGC~1916, the metallicity measurements returned by \textsc{Spexxy} are a key diagnostic. The bulk of the iron lines that allow \textsc{Spexxy} to constrain ${\rm [M/H}]$ are located in the blue part of the MUSE spectral range. As the quality of the AO correction gradually decreases with decreasing wavelength, this part of the spectral range typically has the lowest S/N, especially when the NFM is used.

To independently verify the \texttt{Spexxy} metallicities derived from the MUSE spectra, we make use of the \ion{Ca}{II} triplet lines, which are conveniently located in the red part of the MUSE spectral range where the AO correction works best and hence the S/N is maximised. The \ion{Ca}{II} triplet is commonly used as an iron abundance indicator for old stellar populations \citep[e.g.,][]{2013MNRAS.434.1681C}, and has been used successfully as a diagnostic in MUSE globular cluster observations by \citet{2019ApJ...886...57A, 2020A&A...635A.114H}. For stars on the red giant branch, different calibrations exist in order to convert the equivalent widths (EW) of the \ion{Ca}{II} lines into a metallicity. In this work, we adopt the calibration presented by \citet{2013MNRAS.434.1681C} with the analytical form
$$
{\rm [Fe/H]}=a+b\times{\rm Mag}+c\times \Sigma{\rm Ca}+d\times \Sigma{\rm Ca}^{-1.5}+e\times\Sigma{\rm Ca}\times{\rm Mag}.
$$
Here, $a = -3.45$, $b = 0.11$, $c = 0.44$, $d = -0.65$, $e = 0.03$ according to \citet{2013MNRAS.434.1681C}, $\Sigma{\rm Ca}$ corresponds to the sum of the equivalent width ($\Sigma \rm{EW}$) of the three \ion{Ca}{II} lines measured in ${\text \AA}$, while ${\rm Mag}$ is a luminosity indicator, in our case $V-V_{\rm HB}$, where $V$ is a star's magnitude in the $\rm F606W$ filter and $V_{\rm HB} = 19$ is an average magnitude measured for HB stars within the cluster. In the case of NGC~1916, the magnitude of the horizontal branch is poorly defined, making it difficult to measure this value precisely. Nevertheless, using an underestimated (overestimated) value would yield higher (lower) iron abundances for all stars in our sample almost uniformly, and therefore does not affect the relative ${\rm [Fe/H]}$ values.

The spectra from the cool subsample (with (F606W-F814W)~$>0.6$) were limited to the wavelength range $8400$–$8800~{\text \AA}$, and a three-Gaussian fit was applied to the lines with initial central wavelengths at $\lambda_1 = 8505~{\text \AA}$, $\lambda_2 = 8550~{\text \AA}$, and $\lambda_3 = 8670~{\text \AA}$ using the \texttt{specutils} package\footnote{\url{https://specutils.readthedocs.io/en/stable/index.html}}. We note that the starting values for the line centroids have been Doppler shifted from their restframe values to the expected line-of-sight velocity of NGC~1916 of 265~km/s (cf. Section~\ref{sec:kinematics}).

The [Fe/H] abundances derived from the \ion{Ca}{II} triplet are compared to the \texttt{spexxy} [M/H] estimates in Fig.~\ref{fig:mefe}. While we observe a global shift between the two scales of $\sim0.5$~dex, the shape of the metallicity distribution, in particular the bimodality between a metal-poor and a metal-rich group, is preserved.

\begin{figure}
    \centering
	\includegraphics[width=1.\columnwidth]{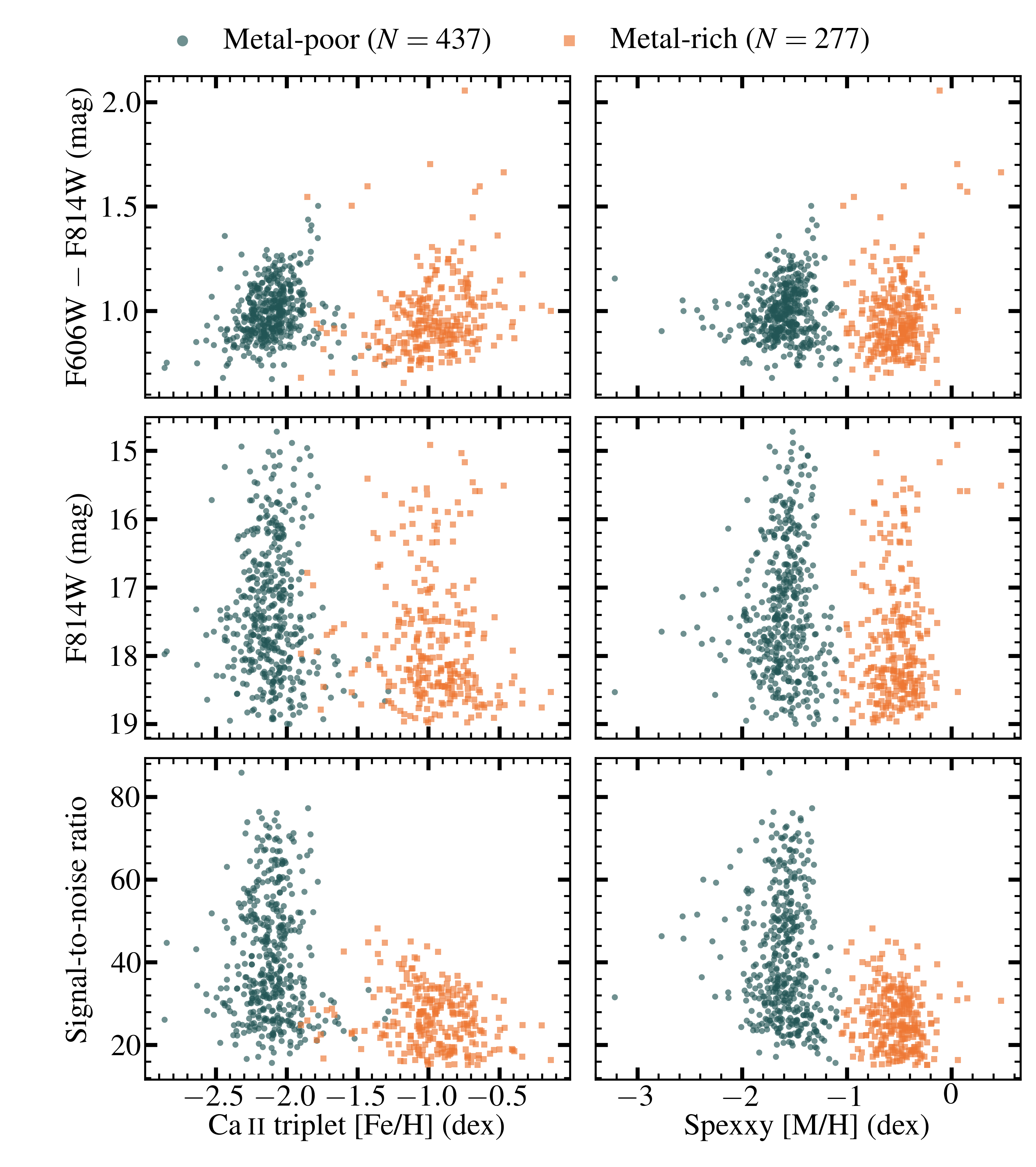}
    \caption{Metallicity measurements from the Ca~II triplet (left) and through full spectrum fitting (right) as a function of colour (top), magnitude (centre), and signal-to-noise ratio (bottom). The comparison includes 714 stars of the cool MUSE subsample. No systematic trends with any of the stars' properties are imminent.}
    \label{fig:test}
\end{figure}

Furthermore, we checked the measured metallicities for biases with respect to photometric colour, magnitude, and signal-to-noise ratio. Fig.~\ref{fig:test} confirms that our metallicity measurements are robust across the parameter ranges relevant to this study. Hereafter, we adopt the \texttt{spexxy} ${\rm [M/H]}$ values. We note, however, that our conclusions do not change qualitatively if the ${\rm [Fe/H]}$ values measured from the \ion{Ca}{II} triplet are used instead.

\section{Kinematics fitting with MCMC}
\label{sec:corner}

\begin{figure}
    \centering
	\includegraphics[width=0.9\columnwidth]{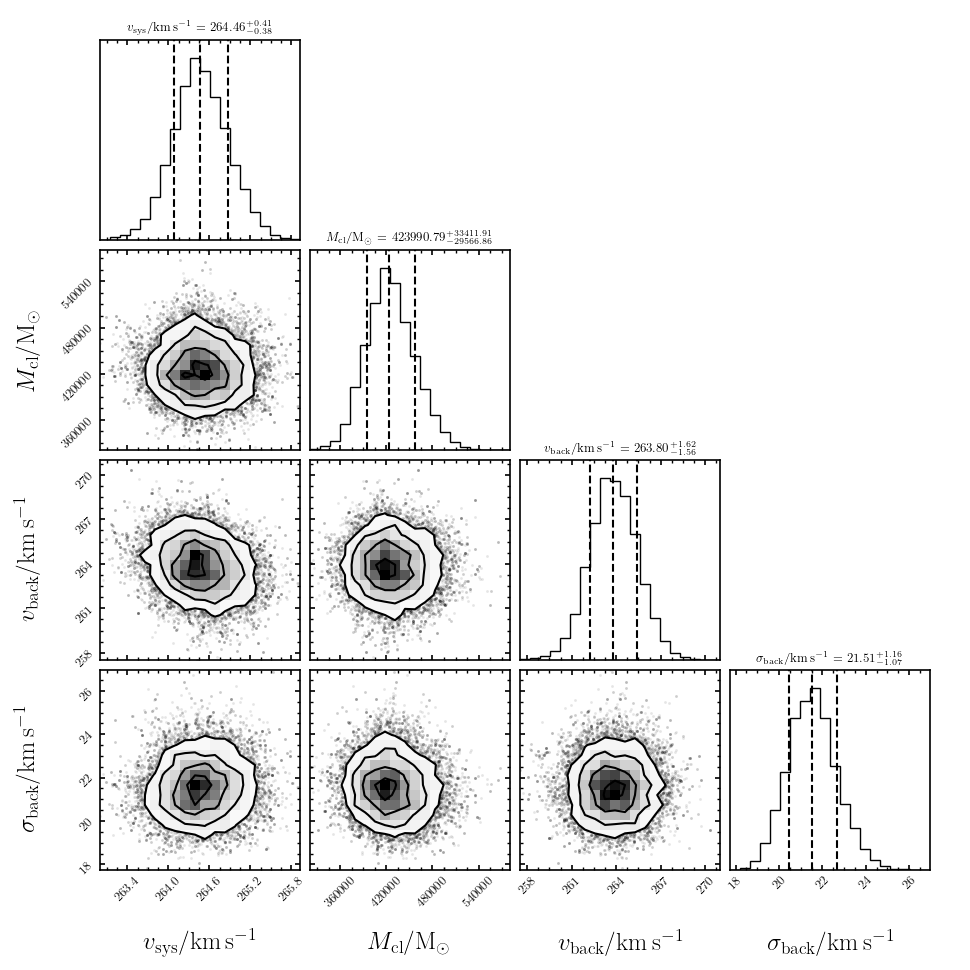}
    \caption{Results of MCMC kinematic fitting presented as a corner plot. The method, values, and uncertainties are described in Sec.~\ref{sec:kinematics}. }
    \label{fig:corner}
\end{figure}

In Fig.~\ref{fig:corner}, we show the fitted parameters from our analysis of the cluster kinematics in a corner plot.


\vspace{-\topsep}


\bsp	
\label{lastpage}
\end{document}